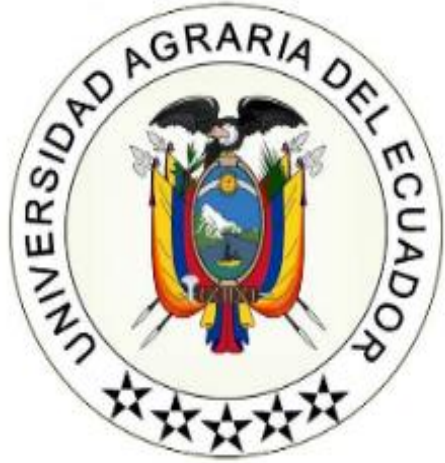

# UNIVERSIDAD AGRARIA DEL ECUADOR

# FACULTAD DE ECONOMÍA AGRÍCOLA

# CARRERA DE ECONOMÍA

**TEMA:**

ANÁLISIS DE LA INCIDENCIA DE LA INVERSIÓN EXTRANJERA DIRECTA
Y LA INVERSIÓN NACIONAL, EN EL CRECIMIENTO ECONÓMICO DE CHILE

**INTEGRANTES:**

ALVEAR KATHERINE

CAMPOZANO JENNER

DURAN PAULETTE

HOLGUIN ROGER

FERNANDO MEJÍA

**TRIUNFO-ECUADOR**

**2023-2024**

# ÍNDICE DE CONTENIDOS

# RESUMEN

Dentro de la presente investigación, que tiene como objetivo determinar la incidencia de la inversión extranjera directa y la inversión nacional en el crecimiento económico de Chile. Permite conocer la relación entre la inversión extranjera directa y la inversión nacional, y como ambas contribuyen al desarrollo económico del país, estos hallazgos pueden proporcionar información valiosa para la formulación de políticas económicas y futuras inversiones. El impacto de la presente investigación es significativo ya que los resultados obtenidos pueden influir en la percepción internacional de Chile como destino de inversión, afectando su posicionamiento en el escenario económico global, si se demuestra que la IED es un motor importante para el crecimiento económico, esto podría aumentar la confianza de los inversores extranjeros. La importancia del proyecto radica en su contribución al conocimiento económico y su capacidad para informar decisiones estratégicas que favorezcan al crecimiento económico sostenible en Chile al entender como la combinación de IED y la inversión nacional afecta el crecimiento económico, se puede buscar un equilibrio que permita un desarrollo económico más estable y evite problemas como la dependencia excesiva de la inversión extranjera o la fuga de capitales; por este motivo se destaca la teoría de la internacionalización que es ampliamente reconocida y utilizada en el estudio de la inversión extranjera directa (IED) ya que proporciona un marco conceptual para entender los motivos y estrategias detrás de las decisiones de las empresas multinacionales para invertir en el extranjero. Por medio de la información recopilada de fuentes como el Banco Central de Chile se analiza el comportamiento de las variables que son el Crecimiento económico de Chile (PIB) como variable dependiente, la inversión extranjera directa como la primera variable dependiente y la inversión nacional como segunda variable independiente. Posteriormente en base al tema se plantea que si existe una relación significativa de causalidad a largo plazo entre la Inversión Extranjera Directa (IED) y entre la Inversión Nacional (IN) en el Crecimiento Económico (PIB) de Chile, siendo esta la hipótesis de la investigación. Para que la información sea clara y concisa se utilizó herramientas para un eficaz análisis de estudio, las cuales fueros a través de fuentes secundarias; para el análisis estadístico, se aplicó procesos de análisis utilizando la herramienta del software Eviews 6 lo cual sirvió para el resultado final de la investigación, en donde se demostró que, para impulsar el crecimiento económico sostenible en Chile, es necesario seguir atrayendo inversiones extranjeras y fomentar la inversión interna.

# ABSTRACT

Within the scope of this research, whose objective is to determine the impact of foreign direct investment and domestic investment on Chile's economic growth, it allows us to understand the relationship between foreign direct investment and domestic investment, and how both contribute to the country's economic development. These findings can provide valuable insights for the formulation of economic policies and future investments. The significance of this research lies in the fact that the obtained results can influence the international perception of Chile as an investment destination, affecting its positioning in the global economic landscape. If it is demonstrated that FDI is a significant driver of economic growth, this could enhance the confidence of foreign investors. The importance of this project is rooted in its contribution to economic knowledge and its ability to inform strategic decisions that favor sustainable economic growth in Chile. By understanding how the combination of FDI and domestic investment impacts economic growth, a balance can be sought that allows for more stable economic development and avoids issues such as excessive reliance on foreign investment or capital flight. For this reason, the theory of internationalization is highlighted, which is widely recognized and used in the study of foreign direct investment (FDI), as it provides a conceptual framework for understanding the motives and strategies behind multinational companies' decisions to invest abroad. Through information collected from sources such as the Central Bank of Chile, the behavior of variables such as Chile's economic growth (GDP) as the dependent variable, foreign direct investment as the first independent variable, and domestic investment as the second independent variable, is analyzed. Subsequently, based on the topic, it is proposed that if there exists a significant long-term causal relationship between Foreign Direct Investment (FDI) and National Investment (NI) in Chile's Economic Growth (GDP), this being the research hypothesis. To ensure clear and concise information, tools were used for effective study analysis, primarily through secondary sources. For statistical analysis, analysis processes were applied using the Eviews 6 software tool, which served for the final outcome of the research, demonstrating that attracting foreign investments and promoting internal investment are necessary to drive sustainable economic growth in Chile.

# INTRODUCCIÓN

El crecimiento económico de un país es importante ya que es un indicador clave que determina el progreso y desarrollo de una nación en el caso de Chile uno de los países más desarrollados de América Latina y con una de las economías más sólidas, el crecimiento es constante y progresivo, con caídas mínimas.

Hay que mencionar que al igual que el resto del mundo Chile tuvo un gran golpe por la pandemia del 2020 repercutiendo significativamente en su economía. "La pandemia de enfermedad por coronavirus provocó en 2020 la mayor contracción del PIB en casi 40 años, del 5,8%, como consecuencia de choques simultáneos en la demanda interna, la demanda externa y la oferta" (CEPAL, 2021).

Uno de los fuertes del crecimiento económico de Chile es la inversión por esto es importante el estudio de la incidencia de la inversión extranjera directa (IED) y la inversión nacional como aspectos claves en el crecimiento económico del país, ya que estas inversiones reflejan un punto fuerte en la economía, el problema específico que aborda la investigación es la competitividad de la IED frente a la inversión nacional.

Comprender cómo la IED interactúa con la inversión nacional es esencial para determinar qué políticas pueden fomentar un crecimiento económico sostenible y equitativo, esta investigación ayudará a evaluar la competitividad de las empresas locales frente a las extranjeras y su capacidad para aprovechar las oportunidades que trae consigo la IED

Como consecuencia, la presente investigación persigue como objetivo principal Determinar la incidencia de la inversión extranjera directa y la inversión nacional en el crecimiento económico de Chile, buscando comprobar o negar, la teoría de internacionalización, partiendo de la hipótesis de que existe una correlación positiva entre las variables independientes y la variable dependiente, esto se determinará mediante un análisis de función de transferencia, modelos dinámicos del método de Fisher y el análisis de vectores autorregresivos.

Se pretende aportar evidencia de tipo empírica sobre la incidencia de estas dos variables sobre el crecimiento económico de Chile en el periodo que abarca la investigación que va desde el 2008 al 2022 con datos trimestrales obtenidos de fuentes confiables, con el fin de reflejar el impacto de las variables en la economía de un país como Chile que tiene una reconocida trayectoria en el desarrollo, con un fuerte dinamismo económico.

## Planteamiento del problema

En el contexto de la economía chilena, la inversión extranjera directa (IED) y la inversión nacional son dos componentes fundamentales que influyen en el crecimiento económico del país. La IED representa los flujos de capital que provienen de inversores extranjeros y se destinan a actividades productivas en Chile, mientras que la inversión nacional abarca los recursos financieros que los inversionistas chilenos destinan a proyectos y negocios dentro del país.

A pesar de su importancia, es necesario analizar en profundidad cómo estas formas de inversión impactan en el crecimiento económico de Chile y si existe una relación diferenciada entre ambas. Es crucial comprender qué factores y sectores son más receptivos a cada tipo de inversión y cómo estas decisiones de inversión influyen en la generación de empleo, el desarrollo de infraestructuras, la productividad y el crecimiento del Producto Interno Bruto (PIB) en el país.

Por lo tanto, el planteamiento del problema de esta investigación es el siguiente:

¿Cuál es la incidencia de la inversión extranjera directa y la inversión nacional en el crecimiento económico de Chile, considerando sus efectos en el empleo, la productividad y el desarrollo de sectores clave el periodo analizado?

# IMPORTANCIA DEL TEMA

El presente estudio sobre la incidencia de la Inversión Extranjera Directa (IED) y la Inversión Nacional (IN) en el crecimiento económico de Chile reviste una importancia fundamental para diversos actores.

En primer lugar, para el Gobierno y formuladores de políticas, los resultados de esta investigación proporcionarán información crucial para diseñar estrategias que impulsen el crecimiento económico sostenible mediante la atracción de inversiones extranjeras y el fomento de la inversión nacional en sectores clave. Además, los inversionistas y empresarios interesados en el mercado chileno podrán tomar decisiones informadas basadas en los datos y conclusiones obtenidas, lo que contribuirá a estimular la inversión privada y el desarrollo económico.

Desde una perspectiva académica, esta investigación enriquecerá la literatura económica existente, al ofrecer nuevos conocimientos y perspectivas sobre la relación entre la IED, la IN y el crecimiento económico en el contexto de Chile. Asimismo, la comprensión de cómo la IED y la IN han afectado el bienestar social y la distribución del ingreso en Chile permitirá formular políticas que fomenten un crecimiento inclusivo y reduzcan las disparidades económicas y sociales.

Por último, los resultados de esta investigación pueden proyectar lecciones para otros países en vías de desarrollo que buscan estimular su crecimiento económico a través de la inversión extranjera y nacional.

## Análisis de la teoría económica

Existe un enfoque muy importante en relación con la inversión extranjera directa y la inversión nacional en el crecimiento económico de Chile: el cual es, la Teoría de la internalización (teoría OLI). Esta teoría es ampliamente reconocida y utilizada en el estudio de la inversión extranjera directa (IED) y proporciona un marco conceptual para entender los motivos y estrategias detrás de las decisiones de las empresas multinacionales para invertir en el extranjero. (Dunning)

## Teoría de la Internalización (OLI)

Olaya y Armijos (2017) sostiene que la inversión extranjera tiene un efecto positivo en el crecimiento económico de las naciones y tiene una relación tanto en el corto como largo plazo. Además, la relación de la IED con el crecimiento económico no se limita a resultados inmediatos. Los efectos de la IED pueden prolongarse en el tiempo, la presencia continua de empresas extranjeras puede fomentar la competitividad, estimular la innovación local y mejorar la calidad y eficiencia de la producción nacional.

De acuerdo con el estudio realizado por Loja y Torres (2013). Según sus hallazgos, la interacción entre la IED y el capital humano en países con niveles muy bajos de este último muestra un efecto directo negativo de la IED. Estos resultados sugieren que Chile presenta una capacidad de absorción de conocimientos tecnológicos transferidos por empresas extranjeras que se considera débil. A diferencia de los enfoques utilizados en otras economías, en nuestro país, se ha adoptado una metodología que emplea variables per cápita para examinar de manera más precisa la relación entre las variables tanto a corto como a largo plazo, lo cual facilita una simplificación del análisis.

Por su parte, Romero (2012) indica que la IED juega un papel complementario, pero no central, en el proceso de crecimiento económico. Él argumenta que la principal alternativa para fomentar el crecimiento es la acumulación de factores, destacando que la acumulación de capital privado nacional es la que produce los mayores beneficios para el país.

En su investigación titulada "La inversión extranjera directa en el Perú y sus implicancias en el crecimiento económico 2009-2015", Bustamante (2017) llega a la conclusión de que la inversión extranjera directa (IED) juega un papel indiscutible como generador de crecimiento en sectores estratégicos de la economía nacional y global. Además, el estudio destaca el destacado desempeño de Perú como economía emergente y atractiva para la atracción positiva de IED en comparación con otros países de la región. Los hallazgos también respaldan empíricamente que un aumento en la IED se traduce en un incremento en la tasa de crecimiento del Producto Interno Bruto (PIB).

# METODOLOGÍA

**Diseño de la investigación**

El análisis de la incidencia de la inversión extranjera directa y la inversión nacional, en el crecimiento económico de chile, es una investigación de carácter formal, fundamentada en la teoría económica de la internalización (OLI), la información fue recopilada del Banco Central de Chile, a través de tablas que contienen las cifras numéricas en miles de millones de dólares americanos, de las variables de estudio.

Para la recolección y análisis de datos correspondiente a las variables establecidas para el desarrollo del proyecto se emplearán los siguientes instrumentos con una metodología econométrica para un eficaz análisis de la información:

➤ Desarrollar de un modelo de regresión lineal empleando el método de mínimos cuadrados ordinarios (MCO).
➤ Estimar un modelo VAR.
➤ Determinar el grado de causalidad de las variables independientes frente a la variable dependiente empleando el método de causalidad de Granger.
➤ Establecer un análisis de impulso-respuesta entre las variables de estudio.
➤ El software Eviews-12, se empleará para efectuar los análisis econométricos a los datos que pertenecen a series de tiempo, datos de corte transversal y datos longitudinales.

**Objetivos**

**Objetivo General**

Determinar la incidencia de la inversión extranjera directa y la inversión nacional en el crecimiento económico de Chile.

**Objetivos específicos**

- Evaluar la incidencia de la inversión extranjera directa y la inversión nacional en el crecimiento económico de Chile mediante un análisis de función de transferencia.
- Evaluar la incidencia de la inversión extranjera directa y la inversión nacional en el crecimiento económico de Chile por modelos dinámicos del método de Fisher.
- Evaluar la incidencia de la inversión extranjera directa y la inversión nacional en el crecimiento económico de Chile mediante vectores autorregresivos.

**Alcance**

La presente investigación evalúa, los datos de series temporales, por motivo de fines académicos, la base de datos abarca desde el año 2008 hasta el 2023 las cuentas nacionales de Chile de forma trimestral.

**Fuente de información**

La selección de información para estimar la incidencia de la inversión extranjera directa y la inversión nacional en el crecimiento económico de Chile de manera trimestral desde el 2008 al 2023 fue obtenida de la base de datos del Banco central de Chile, los datos escogidos presentan características de confiabilidad y validez.

**Tratamiento de datos**

Se escrutará la base de datos tomada del Banco Central de las cuentas nacionales chilenas trimestrales, tanto para la variable dependiente el Crecimiento económico medida por el PIB, y las variables independientes la Inversión Extranjera Directa y la Inversión Nacional del año 2008 al 2023 en la que posterior se empelará una metodología de tipo correlacional.

Posteriormente se ingresa al software econométrico Eviews 6, los datos de las variables, se obtuvieron 61 observaciones al ingresar los datos al programa, para de esta forma realizar el modelo de regresión lineal, donde se determinará la ecuación correspondiente a la regresión lineal y las ecuaciones para cada una de las variables en el modelo VAR, además de los resultados al aplicar el modelo de causalidad de Granger y la función impulso-respuesta.

Luego del análisis e interpretación se realizará la verificación de la hipótesis planteada, los resultados se interpretarán mediante un análisis de tipo descriptivo.

**Hipótesis**

Se plantea la siguiente hipótesis para el presente proyecto, que empleará un modelo econométrico VAR, aplicando la causalidad de Granger y la función de impulso-respuesta:

Existe una relación significativa de causalidad a largo plazo entre la Inversión Extranjera Directa (IED) y entre la Inversión Nacional (IN) en el Crecimiento Económico (PIB) de Chile.

**Definición de variables**

**Primera variable independiente**

Inversión extranjera directa: se expresa como el registro de inversiones de una entidad residente de una economía (inversionista directo), en otra economía o país teniendo diferentes objetivos y funciones. Tiene como finalidad influir significativamente en la gestión de su inversión.

**Técnica de tratamiento de información**

➢ Fuentes secundarias.
➢ Uso de estadística descriptiva.
➢ Empleo de software estadístico Eviews 6.

**Resultados esperados**

Determinar si la variable independiente Inversión extranjera directa causa en el sentido de Granger a la variable dependiente PIB, así también conocer si los shocks ocaurridos a lo largo del tiempo en la variable Inversión extranjera directa, causan respuestas en la variable dependiente.

**Segunda variable independiente**

Inversión nacional: se refiere al gasto en bienes de capital, infraestructura y otros activos productivos realizados por empresas y ciudadanos chilenos dentro del territorio del país, lo cual será medido a través de la FBK

**Técnica de tratamiento de información**

➢ Fuentes secundarias.
➢ Uso de estadística descriptiva.
➢ Empleo de software estadístico Eviews 6.

**Resultados esperados**

Determinar si la variable independiente Inversión nacional causa en el sentido de Granger a la variable dependiente PIB, así también conocer si los shocks ocurridos a lo largo del tiempo en la variable Inversión nacional, causan respuestas en la variable dependiente.

**Variable dependiente**

Crecimiento económico (PIB): es un indicador que mide el aumento sostenido y constante de la producción de bienes y servicios de una economía en un período determinado, es uno de los principales objetivos de las políticas económicas de los países ya que se considera un indicador clave del progreso y desarrollo de una nación. Será obtenido mediante el PIB de Chile del periodo establecido.

**Técnica de tratamiento de información**

➢ Fuentes secundarias.
➢ Uso de estadística descriptiva.
➢ Empleo de software estadístico Eviews 6.

**Resultados esperados**

Determinar si la variable PIB, correspondiente al crecimiento económico de Chile, es causada por las variables inversión nacional y/o inversión extranjera directa, así también conocer cómo responde la variable PIB a los shocks de las variables independientes ya mencionadas.

## DESARROLLO Y DISCUCIÓN

**Primer objetivo:** Evaluar la incidencia de la inversión extranjera directa y la inversión nacional en el crecimiento económico de Chile mediante un análisis de función de transferencia.

**Desarrollo**

### Resultados del modelo

### Gráfico 1. Ecuación de regresión lineal

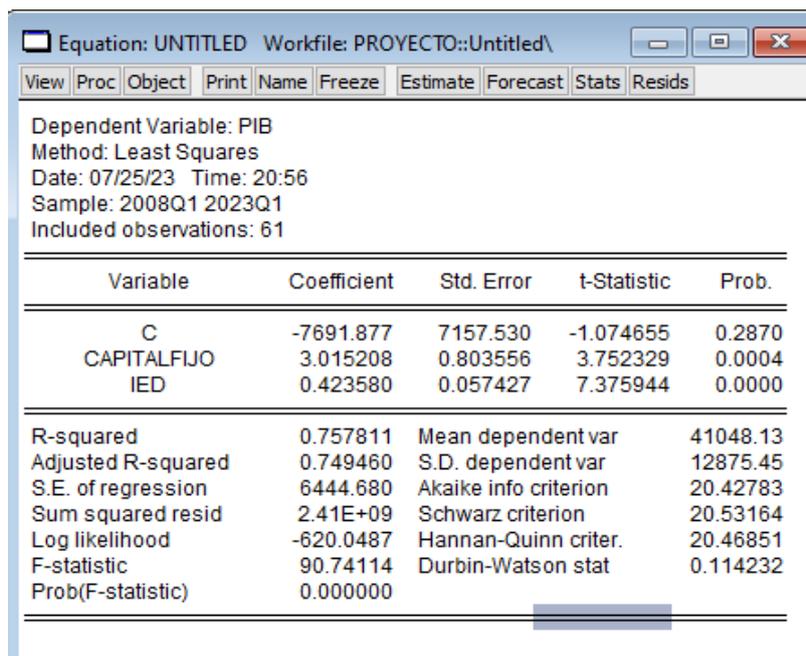

**Fuente: Los Autores, 2023**

**Modelo econométrico**

El planteamiento del modelo econométrico permite analizar de manera estructural la investigación, en otras palabras, la cuantificación de la relación de las variables implicadas, teniendo en consideración los valores de los parámetros estimados, por ello es necesario la especificación de las variables, su ubicación y cómo serán interpretadas. Se describe lo siguiente:

**Variable dependiente:** Crecimiento Económico – Chile (PIB)

**Variable independiente 1:** Inversión Extranjera Directa (IED)

**Variable independiente 2:** Inversión Nacional (CAPITAL FIJO)

En donde la interpretación del modelo está dada de la siguiente manera:

$$Y = B\_0 + B\_1 X\_1 + B\_2 X\_2 + \mu$$

Siendo:

**Y** = Crecimiento económico (PIB) - Chile

**B0**= Intercepto de la regresión lineal múltiple

**X1**= Inversión Extranjera Directa

**X2**= Inversión Nacional

**μ** = Perturbación aleatoria

Dentro de un análisis econométrico es interpretado de la siguiente manera: B0 como el intercepto de la regresión lineal múltiple, B1 como la pendiente de la regresión donde X1 es la variable independiente que es Inversión Extranjera Directa, también X2 que es la segunda variable independiente catalogada como Inversión Nacional y finalmente se interpreta μ como la perturbación aleatoria, es decir, la inclusión de forma conjunta de todos los valores no especificados en el modelo.

En la tabla 1 podemos observar el PIB, considerado como la variable dependiente, mientras que la Inversión Extranjera Directa (IED) y la Inversión Nacional (CAPITAL FIJO) es tomado como variable independiente.

**Interpretación de los coeficientes**

- Al observar los coeficientes, la constante (C), toma el valor de -7691.877, es decir si se presenta la situación en la que no hay Inversión nacional, ni Inversión extranjera directa, (C) se mantendrá en dicho valor constante, al ser negativo, existirá una pérdida.
- Al analizar los resultados de la variable CAPITAL FIJO, tenemos un valor de 3.015, lo que se puede interpretar como que el aumento de cada unidad de dólar que aumente la Inversión nacional, el PIB incrementará un valor en un 3.015.
- La variable IED, muestra un valor de 0.4235, esto nos dice que, por cada unidad de dólar invertido en Inversión extranjera directa, el PIB aumenta en 0.4235.

Otros coeficientes relevantes para determinar la validez de la hipótesis son detallados a continuación:

**Resultados de la interpretación**

**R-squared:** Constituye el grado de explicación de las variables independientes sobre las variables dependientes, se obtuvo un r-squared de 75%, es decir el 75% de la variable dependiente, es predicha por la variable dependiente.

**Adjusted r-squared:** Permite estimar el incremento neto de r cuadrado cuando se integre un nuevo regresor al modelo, para este parámetro se obtuvo el valor de 74%.

**Prob (F-statistic):** Si este coeficiente es mayor al 5%, se argumenta que los coeficientes (bo,b1...bk) son iguales a cero, es decir el modelo no es aplicable. Para el modelo propuesto se obtuvo una probabilidad del estadístico F de 0.00000, por lo que concluimos que el modelo y sus variables son representativas.

**Pruebas individuales**

**Prob (t-Statistic):**

- C: Se obtuvo un estadístico t de 0.2870, es decir mayor al 5%, se concluye con el 95% de confianza, que el término de la constante no es representativo en el modelo.
- CAPITAL FIJO: se obtuvo un estadístico t de 0.0004, por lo que se concluye con un 95% de confianza que el término de la inversión nacional explica al PIB y es representativa para el modelo.
- IED: Se obtuvo un valor del estadístico t de 0.0000, por lo que se concluye con un 95% de confianza que el término de la inversión extranjera directa explica al PIB y es representativa en el modelo.

**Pruebas y supuestos de estimación**

**Pruebas de detección y corrección de estacionariedad mediante el método de Phillips-Perron.**

**Gráfico 2. Test de Phillips-Perron - PIB**

**Fuente: Los Autores, 2023.**

**Gráfico 3. Test de Phillips-Perron – CAPITAL FIJO**

**Fuente: Los Autores, 2023.**

**Gráfico 4. Test de Phillips-Perron – IED**

**Fuente: Los Autores, 2023.**

Se obtiene un p valor mayor que 0.05 en los resultados de las variables PIB, IED, CAPITAL FIJO con el contraste de Phillips Perron, lo que indica ausencia de estacionariedad. Lo que nos lleva a considerar sus primeras diferencias.

**Gráfico 5. Test de Phillips-Perron - PIB (estacionariedad corregida)**

**Fuente: Los Autores, 2023.**

**Gráfico 6. Test de Phillips-Perron – CAPITAL FIJO (estacionariedad corregida)**

**Fuente: Los Autores, 2023.**

**Gráfico 7. Test de Phillips-Perron – IED (estacionariedad corregida)**

**Fuente: Los Autores, 2023.**

Se obtiene un p-valor menor que 0,05 en los resultados del contraste de Phillips Perron, lo que indica estacionariedad en la primera diferencia de las variables PIB, CAPITAL FIJO, IED.

**Test y corrección de cointegración de los residuos de las series mediante el método de Phillips-Perron.**

Realizamos el test de cointegración de Phillips-Perron, para una copia de los residuos de la regresión lineal estimada.

**Gráfico 8. Test de Phillips-Perron – Resid01**

**Fuente: Los Autores, 2023.**

Se obtiene un p-valor mayor que 0.05 en los resultados del contraste de Phillips Perron. Llegamos entonces a la conclusión de que las variables del modelo no cointegran y este puede ser espurio.

Aplicamos las correcciones necesarias que constan en anexos.

**Gráfico 9. Test de Phillips-Perron – Resid02 (Corregido)**

**Fuente: Los Autores, 2023.**

Se obtiene un valor menor que 0.05 en el resultado de Phillips-Perron. Se concluye que las variables del nuevo modelo cointegran y no es espurio.

La cointegración estimada será:

$$Y = 3442.796 + 2.151688\,CF + 0.300825\,IED - 2527.361\,D1 + 6773.382\,D2 + 13359.80\,D3$$

**Detección de heterocedasticidad**

**Prueba de Glejser**

Se escoge el supuesto para verificar la ausencia o presencia de heterocedasticidad en nuestro modelo mediante el test de Glejser.

**Gráfico 10. Test de Glejser**

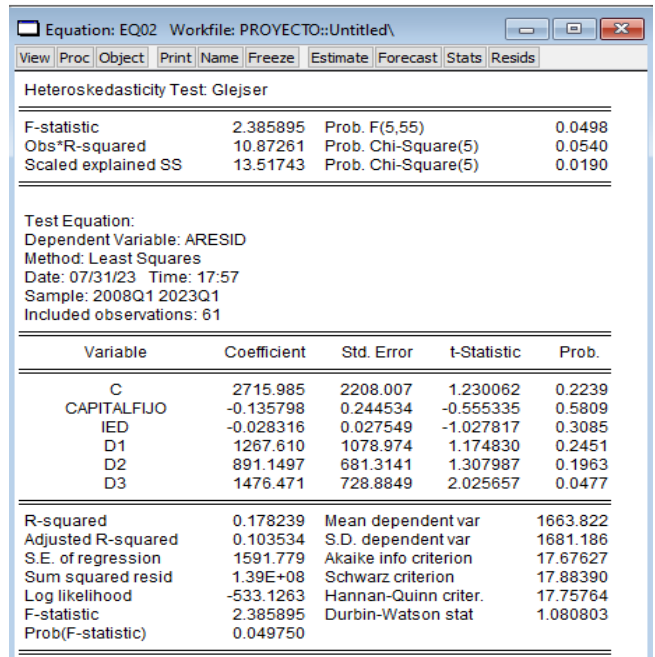

**Fuente: Los Autores, 2023.**

La probabilidad de F es menor al 5%, por ende, se confirma que el modelo no tiene problemas de heterocedasticidad, por ende, no es necesaria su corrección

**Test de normalidad**

Se verifica la ausencia o presencia de normalidad en nuestro modelo mediante el análisis de Jarque-Bera.

**Gráfico 11. Test de Jarque-Bera**

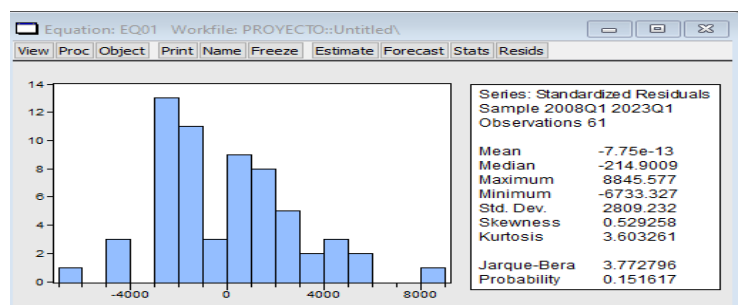

**Fuente: Los Autores, 2023.**

El resultado de la probabilidad de Jarque-Bera en la tabla es 0.1516 y la kurtosis se encuentra por fuera del rango de -2 y 2.

En el caso de nuestro modelo, la probabilidad de Jarque-Bera es mayor al 5%, por lo tanto, hay normalidad en los residuos.

**Test de linealidad**

Se verifica la ausencia o presencia de linealidad en nuestro modelo mediante el análisis de Ramsey RESET.

**Gráfico 12. Test de Ramsey Reset**

**Fuente: Los Autores, 2023.**

Si la probabilidad es menor al 5% existe problema de linealidad y al ser mayor al 5%, no existe.

**Test de autocorrelación**

Se verifica la ausencia o presencia de autocorrelación en nuestro modelo mediante el análisis de estadístico Durbin-Watson.

**Gráfico 13. Estadístico Durbin-Watson**

Podemos visualizar que DW se encuentra en 1,27%. Por ende, podemos afirmar si hay problema de autocorrelación en nuestro modelo.

Se lo corrige aplicando un rezago en el modelo con la variable Ar (1)

**Gráfico 14. Corrección de Autocorrelación**

Dependent Variable: PIB
Method: Least Squares
Date: 07/31/23   Time: 18:03
Sample (adjusted): 2008Q2 2023Q1
Included observations: 60 after adjustments
Convergence achieved after 158 iterations

| Variable | Coefficient | Std. Error | t-Statistic | Prob. |
|---|---|---|---|---|
| C | 2768061. | 1.28E+08 | 0.021627 | 0.9828 |
| CAPITALFIJO | 2.494273 | 0.227443 | 10.96658 | 0.0000 |
| IED | -0.315447 | 0.126999 | -2.483854 | 0.0162 |
| D1 | 1220.924 | 1623.091 | 0.752221 | 0.4552 |
| D2 | 256.7456 | 1620.019 | 0.158483 | 0.8747 |
| D3 | -1589.958 | 1608.821 | -0.988275 | 0.3275 |
| AR(1) | 0.999688 | 0.014566 | 68.63214 | 0.0000 |
| R-squared | 0.986465 | Mean dependent var | | 41330.79 |
| Adjusted R-squared | 0.984932 | S.D. dependent var | | 12791.82 |
| S.E. of regression | 1570.195 | Akaike info criterion | | 17.66507 |
| Sum squared resid | 1.31E+08 | Schwarz criterion | | 17.90941 |
| Log likelihood | -522.9520 | Hannan-Quinn criter. | | 17.76064 |
| F-statistic | 643.7835 | Durbin-Watson stat | | 2.084228 |
| Prob(F-statistic) | 0.000000 | | | |

| | | |
|---|---|---|
| Inverted AR Roots | 1.00 | |

Podemos visualizar que DW se encuentra en 2,84%. Por ende, podemos afirmar que se ha corregido el problema de autocorrelación en nuestro modelo y este es el mejor modelo propuesto debido a que no presenta ningún problema que lo pueda considerar como espurio.

**Segundo objetivo:** Evaluar la incidencia de la inversión extranjera directa y la inversión nacional en el crecimiento económico de Chile por modelos dinámicos del método de Fisher.

**Desarrollo**

**Resultados del modelo**

**Gráfico 15. Ecuación de regresión lineal (retardo dinámico de Fisher)**

**Fuente: Los Autores, 2023.**

**Modelo econométrico**

**Variable dependiente:** Crecimiento Económico – Chile (PIB)

**Variable independiente 1:** Inversión Extranjera Directa (ZIED)

**Variable independiente 2:** Inversión Nacional (ZCF)

En donde la interpretación del modelo está dada de la siguiente manera:

$$Y= B_0 + B_1 X_1 + B_2 X_2 + \mu$$

Siendo:

**Y** = Crecimiento económico (PIB) - Chile

**B0**= Intercepto de la regresión lineal múltiple

**X1**= Inversión Extranjera Directa

**X2**= Inversión Nacional

**μ** = Perturbación aleatoria

**Interpretación de los coeficientes**

- Al observar los coeficientes, la constante (C), toma el valor de -6285.040, es decir si se presenta la situación en la que no hay Inversión nacional, ni Inversión extranjera directa, (C) se mantendrá en dicho valor constante, al ser negativo, existirá una pérdida.
- Al analizar los resultados de la variable ZCF, tenemos un valor de 0.08190, lo que se puede interpretar como que el aumento de cada unidad de dólar que aumente la Inversión nacional, el PIB incrementará un valor en un 0.08190.
- La variable ZIED, muestra un valor de 0.0120, esto nos dice que, por cada unidad de dólar invertido en Inversión extranjera directa, el PIB aumenta en 0.0120.

Otros coeficientes relevantes para determinar la validez de la hipótesis son detallados a continuación:

**Resultados de la interpretación**

**R-squared:** Constituye el grado de explicación de las variables independientes sobre las variables dependientes, se obtuvo un r-squared de 69%, es decir el 69% de la variable dependiente, es predicha por la variable dependiente.

**Adjusted r-squared:** Permite estimar el incremento neto de r cuadrado cuando se integre un nuevo regresor al modelo, para este parámetro se obtuvo el valor de 68%.

**Prob (F-statistic):** Si este coeficiente es mayor al 5%, se argumenta que los coeficientes (bo,b1…bk) son iguales a cero, es decir el modelo no es aplicable. Para el modelo propuesto se obtuvo una probabilidad del estadístico F de 0.00000, por lo que concluimos que el modelo y sus variables son representativas.

**Pruebas individuales**

**Prob (t-Statistic):**

- C: Se obtuvo un estadístico t de 0.5834, es decir mayor al 5%, se concluye con el 95% de confianza, que el término de la constante no es representativo en el modelo.
- ZCF: se obtuvo un estadístico t de 0.0224, menor al 5%, por lo que se concluye con un 95% de confianza que el término de la inversión nacional explica al PIB y es representativa en para el modelo.
- ZIED: Se obtuvo un valor del estadístico t de 0.0000, por lo que se concluye con un 95% de confianza que el término de la inversión extranjera directa explica al PIB y es representativa en el modelo.

**Pruebas y supuestos de estimación**

**Estadístico Durbin Watson**

**Gráfico 16. Autocorrelación Durbin-Watson**

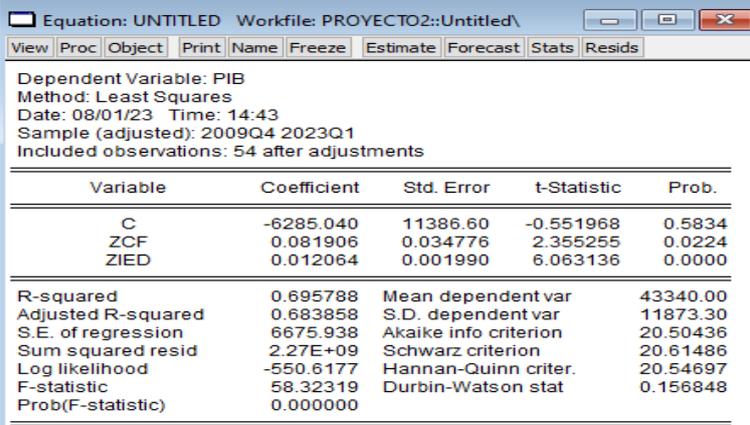

**Fuente: Los Autores, 2023.**

El ajuste del modelo es bueno, salvo el problema de la autocorrelación residual derivada del valor tan bajo del estadístico de Durbin Watson.

## Gráfico 17. Autocorrelación Correjida

**Fuente: Los Autores, 2023.**

Se ajusta el modelo con una estructura AR (2) en sus residuos. Se obtienen los resultados. El modelo presenta buena significatividad individual y conjunta de los parámetros estimados, altos coeficientes de determinación y un estadístico de Durbin Watson casi igual a 2, el ajuste es correcto.

## Detección de heterocedasticidad

## Prueba de Glejser

### Gráfico 17. Test de Heterocedasticidad

**Fuente: Los Autores, 2023.**

La probabilidad de F es mayor al 5%, por ende, se confirma que el modelo no tiene problemas de heterocedasticidad, por ende, no es necesaria su corrección

**Test de Normalidad**

Se verifica la ausencia o presencia de normalidad en nuestro modelo mediante el análisis de Jarque-Bera.

**Gráfico 18. Prueba de Normalidad Jarque-Bera**

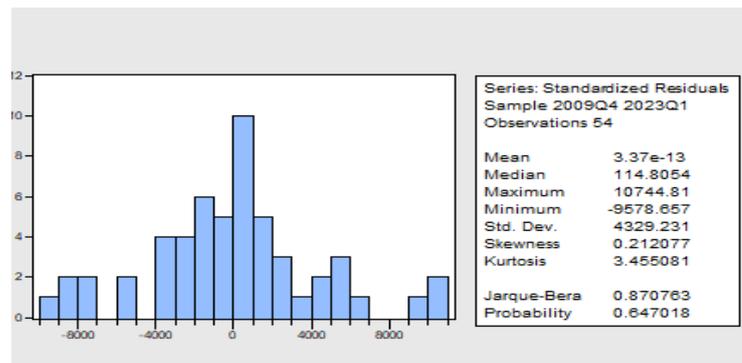

**Fuente: Los Autores, 2023.**

El resultado de la probabilidad de Jarque-Bera en la tabla es 0.6470 y la kurtosis se encuentra por fuera del rango de -2 y 2.

En el caso de nuestro modelo, la probabilidad de Jarque-Bera es mayor al 5%, por lo tanto, hay normalidad en los residuos.

**Test de Linealidad**

Se verifica la ausencia o presencia de linealidad en nuestro modelo mediante el análisis de Ramsey RESET.

**Gráfico 19. Test de Linealidad Ramsey Reset**

Equation: EQ01    Workfile: PROYECTO2::Untitled\

View  Proc  Object  Print  Name  Freeze  Estimate  Forecast  Stats  Resids

Ramsey RESET Test:

| | Value | | df | Probability |
|---|---|---|---|---|
| F-statistic | 8.354363 | Prob. F(1,46) | | 0.0059 |
| Log likelihood ratio | 8.677941 | Prob. Chi-Square(1) | | 0.0032 |

Test Equation:
Dependent Variable: PIB
Method: Least Squares
Date: 08/01/23   Time: 15:22
Sample: 2010Q2 2023Q1
Included observations: 52
Convergence achieved after 6 iterations

| Variable | Coefficient | Std. Error | t-Statistic | Prob. |
|---|---|---|---|---|
| C | 17083.61 | 3785.404 | 4.513022 | 0.0000 |
| ZCF | 0.006809 | 0.011317 | 0.601652 | 0.5504 |
| ZIED | 0.004238 | 0.000670 | 6.326675 | 0.0000 |
| FITTED^2 | 8.55E-06 | 3.81E-07 | 22.44478 | 0.0000 |
| AR(1) | -0.093481 | 0.149672 | -0.624572 | 0.5353 |
| AR(2) | -0.138354 | 0.153802 | -0.899561 | 0.3730 |

| | | | |
|---|---|---|---|
| R-squared | 0.963459 | Mean dependent var | 44004.71 |
| Adjusted R-squared | 0.959487 | S.D. dependent var | 11590.27 |
| S.E. of regression | 2332.865 | Akaike info criterion | 18.45575 |
| Sum squared resid | 2.50E+08 | Schwarz criterion | 18.68089 |
| Log likelihood | -473.8495 | Hannan-Quinn criter. | 18.54206 |
| F-statistic | 242.5725 | Durbin-Watson stat | 2.051043 |
| Prob(F-statistic) | 0.000000 | | |

| Inverted AR Roots | -.05+.37i | -.05-.37i |
|---|---|---|

El resultado de F en la tabla es 0.0059

La probabilidad del F estadístico es menor al 5%, podemos afirmar que no existe problema de linealidad.

**Test de Autocorrelación**

Se verifica la ausencia o presencia de autocorrelación en nuestro modelo mediante el análisis de estadístico Durbin-Watson.

**Gráfico 20. Test de autocorrelación corregida.**

Podemos visualizar que DW se encuentra en 1,99 %. Por ende, podemos afirmar no hay problema de autocorrelación en nuestro modelo. Obteniendo así, el mejor modelo posible para las series estimadas.

**Tercer objetivo:** Evaluar la incidencia de la inversión extranjera directa y la inversión nacional en el crecimiento económico de Chile mediante vectores autorregresivos.

**Modelo de Vectores Autorregresivos**

**Modelo econométrico**

**Tabla 1. Modelo VAR**

Vector Autoregression Estimates
Date: 08/01/23  Time: 19:47
Sample (adjusted): 2009Q2 2023Q1
Included observations: 56 after adjustments
Standard errors in ( ) & t-statistics in [ ]

|         | PIB      | CF        | IED       |
|---------|----------|-----------|-----------|
| PIB(-1) | 0.469584 | -0.106175 | -0.076451 |
|  | (0.19452) | (0.07307) | (0.20156) |
|  | [ 2.41411] | [-1.45298] | [-0.37930] |
| PIB(-2) | 0.577223 | 0.160100 | 0.367789 |
|  | (0.21624) | (0.08124) | (0.22407) |
|  | [ 2.66930] | [ 1.97079] | [ 1.64138] |
| PIB(-3) | 0.070912 | 0.085277 | 0.197547 |
|  | (0.23923) | (0.08987) | (0.24789) |
|  | [ 0.29643] | [ 0.94890] | [ 0.79693] |
| PIB(-4) | 0.533092 | 0.006404 | -0.541997 |
|  | (0.23621) | (0.08874) | (0.24476) |
|  | [ 2.25686] | [ 0.07217] | [-2.21439] |
| PIB(-5) | -0.524897 | -0.139985 | -0.180093 |
|  | (0.18859) | (0.07085) | (0.19541) |
|  | [-2.78335] | [-1.97592] | [-0.92160] |
| CF(-1) | 0.409645 | 1.005883 | 0.805520 |
|  | (0.48497) | (0.18219) | (0.50252) |
|  | [ 0.84469] | [ 5.52117] | [ 1.60295] |
| CF(-2) | -1.592630 | -0.399548 | -0.527640 |
|  | (0.63898) | (0.24004) | (0.66211) |
|  | [-2.49247] | [-1.66448] | [-0.79691] |
| CF(-3) | -0.619526 | -0.190275 | -0.273190 |
|  | (0.69815) | (0.26227) | (0.72342) |
|  | [-0.88739] | [-0.72548] | [-0.37764] |
| CF(-4) | 0.685952 | 0.592633 | 1.667247 |
|  | (0.70721) | (0.26568) | (0.73281) |
|  | [ 0.96995] | [ 2.23067] | [ 2.27515] |
| CF(-5) | 0.308967 | -0.248007 | 0.102626 |
|  | (0.54832) | (0.20599) | (0.56817) |
|  | [ 0.56348] | [-1.20400] | [ 0.18063] |
| IED(-1) | 0.056520 | -0.031755 | 0.802928 |
|  | (0.15485) | (0.05817) | (0.16045) |
|  | [ 0.36501] | [-0.54589] | [ 5.00418] |
| IED(-2) | 0.127005 | 0.083757 | 0.118787 |
|  | (0.20024) | (0.07523) | (0.20749) |
|  | [ 0.63425] | [ 1.11341] | [ 0.57248] |
| IED(-3) | -0.067690 | 0.004149 | 0.146091 |
|  | (0.20306) | (0.07629) | (0.21042) |
|  | [-0.33334] | [ 0.05439] | [ 0.69430] |
| IED(-4) | -0.118142 | -0.125452 | -0.487628 |
|  | (0.20580) | (0.07731) | (0.21325) |
|  | [-0.57407] | [-1.62267] | [-2.28666] |
| IED(-5) | -0.018828 | 0.070208 | 0.413132 |
|  | (0.15795) | (0.05934) | (0.16367) |
|  | [-0.11920] | [ 1.18318] | [ 2.52414] |
| C | 4878.744 | 2173.394 | -8370.813 |
| | (3009.67) | (1130.64) | (3118.64) |
| | [ 1.62102] | [ 1.92227] | [-2.68413] |
| R-squared | 0.989342 | 0.864282 | 0.994400 |
| Adj. R-squared | 0.985345 | 0.813388 | 0.992300 |
| Sum sq. Resids | 87950011 | 12412151 | 94433491 |
| S.E. equation | 1482.818 | 557.0492 | 1536.502 |
| F-statistic | 247.5269 | 16.98195 | 473.5050 |
| Log likelihood | -478.9345 | -424.1079 | -480.9261 |
| Akaike AIC | 17.67623 | 15.71814 | 17.74736 |
| Schwarz SC | 18.25491 | 16.29681 | 18.32603 |
| Mean dependent | 42621.81 | 10626.30 | 43318.22 |
| S.D. dependent | 12248.71 | 1289.507 | 17509.71 |
| Determinant resid covariance (dof adj.) | 8.64E+17 | | |
| Determinant resid covariance | 3.15E+17 | | |
| Log likelihood | -1366.544 | | |
| Akaike information criterion | 50.51943 | | |
| Schwarz criterion | 52.25544 | | |
| Number of coefficients | 48 | | |

**Fuente: Los Autores, 2023.**

Para la aplicación del modelo VAR se trabajará con las siguientes ecuaciónes:

$$Yt = \alpha_0 + \alpha_1 Y_{t-p} + \alpha_2 X_{t-p} + \alpha_3 Z_{t-p} + \mu t$$

$$PIB = \alpha_0 + \alpha_1 PIB_{t-p} + \alpha_2 IED_{t-p} + \alpha_3 inversion\ nacional_{t-p} + \mu t$$

$$Xt = \beta_0 + \beta_1 X_{t-p} + \beta_2 Y_{t-p} + \beta_3 Z_{t-p} + wt$$

$$IED = \beta_0 + \beta_1 IED_{t-p} + \beta_2 PIB_{t-p} + \beta_3 Inversión\ nacional_{t-p} + wt$$

$$Zt = \lambda_0 + \lambda_1 Z_{t-p} + \lambda_2 X_{t-p} + \lambda_3 Y_{t-p} + \varepsilon t$$

$$In, Nacional = \lambda_0 + \lambda_1 inversión\ nacional_{t-p} + \lambda_2 PIB_{t-p} + \lambda_3 IED_{t-p} + \varepsilon t$$

**Donde:**

**Yt, Xt y Zt:** son vectores con g variables explicativas, representado por el PIB, la IED y la inversión nacional, respectivamente.

$\boldsymbol{\alpha_0, \beta_0\ y\ \lambda_0}$: son el valor de la intersección o el término constante del modelo VAR para cada ecuación.

$\mathbf{IED_{t-p}}$: es el valor rezagado de la variable inversión extrangera directa p periodos anteriores al tiempo t.

$\mathbf{PIB_{t-p}}$: es el valor rezagado de la variable PIB p periodos anteriores al tiempo t.

$\boldsymbol{\lambda_1}$ **inversión** $\mathbf{nacional_{t-p}}$: es el valor rezagado de la variable inversión nacional p periodos anteriores al tiempo t.

$\boldsymbol{\alpha_1, \alpha_2\ y\ \alpha_3}$: son matrices de los coeficientes a estimar en el modelo Xt.

$\boldsymbol{\beta_1, \beta_2\ y\ \beta_3}$: son matrices de los coeficientes a estimar en el modelo Yt.

$\lambda_1, \lambda_2$ y $\lambda_3$: son matrices de los coeficientes a estimar en el modelo Zt.

$\mu t, wt$ y $\varepsilon t$: son el término del error de cada ecuación, respectivamente, y captura la parte no explicada de las variables independientes.

El modelo se puede estimar utilizando mínimos cuadrados ordinarios ya que el estimador MCO es imparcial y consistente.

El orden del modelo VAR(p), se determinará mediante la prueba de rezago óptimo empleando el software estadístico Eviews 6.

**Estabilidad del modelo**

**Gráfico 21. Prueba de estabilidad**

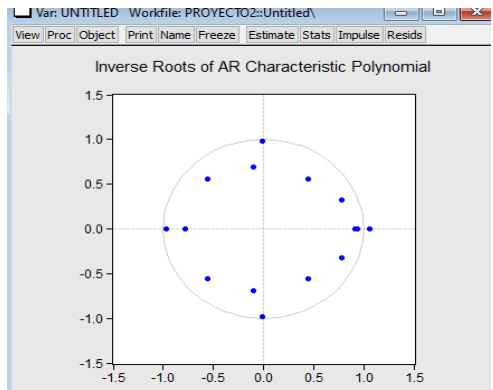

**Fuente: Los Autores, 2023.**

Las raíces inversas del polinomio característico del modelo se encuentran dentro del círculo unitario, por tanto, se confirma la estabilidad del modelo.

**Función impulso-respuesta**

Se generan los nueve gráficos correspondientes a las tres series, sin embargo, para nuestro análisis únicamente nos centraremos en la respuesta del producto interno bruto a las variables CF y IED:

**Gráfico 22. Función Impulso-Respuesta**

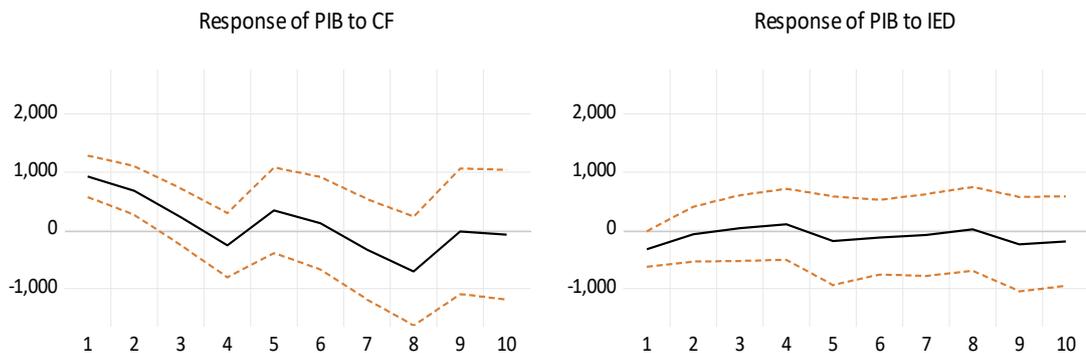

La función de impulso respuesta, muestra el shock de una variable sobre otra a lo largo de su trayectoria, como se evidencia en los gráficos obtenidos, al aplicar la función de impulso-respuesta de Cholesky en eviews 6, tenemos que para el primer gráfico correspondiente a "PIB to CF", los errores estándar de la serie, correspondientes a la línea punteada de color naranja, caen en zona positiva conjuntamente para los primeros dos periodos, sin embargo, este efecto no se mantiene en el tiempo, por lo que podemos inferir que la variable PIB no presenta una respuesta positiva, ni negativa, a los impulsos ocasionados debido a los shocks de la variable CF, a excepción de los dos periodos iniciales, en donde sí se vio influenciada.

Para el gráfico "PIB to IED", se obtuvo que los errores estándar de la serie no caen conjuntamente en zona positiva, tampoco así en zona negativa, por lo que se concluye que la variable PIB no presenta ninguna respuesta a los impulsos ocasionados debido a los shocks ocurridos en la variable IED.

**Descomposición de varianza de Cholesky**

**Gráfico 23. Descomposición de varianza**

| Variance Decomposition of PIB: | | | | |
|---|---|---|---|---|
| Period | S.E. | CF | IED | PIB |
| 1 | 1482.818 | 39.10848 | 4.784022 | 56.10750 |
| 2 | 1714.669 | 45.05529 | 3.732080 | 51.21263 |
| 3 | 1920.416 | 37.35041 | 3.010714 | 59.63887 |
| 4 | 2165.867 | 30.78625 | 2.586950 | 66.62680 |
| 5 | 2806.364 | 19.83525 | 1.964050 | 78.20070 |
| 6 | 2990.284 | 17.63510 | 1.892144 | 80.47276 |
| 7 | 3260.275 | 15.85573 | 1.653352 | 82.49092 |
| 8 | 3563.923 | 17.16124 | 1.386864 | 81.45190 |
| 9 | 4005.390 | 13.58907 | 1.449915 | 84.96102 |
| 10 | 4168.541 | 12.57872 | 1.540720 | 85.88056 |

Cholesky Ordering: CF IED PIB

Para el primer trimestre, tenemos que la variable PIB es impulsada en un 39% por la variable CF y un 4,7% por la variable IED, así mismo para el trimestre dos, tenemos que la varianza de va variable PIB viene dada en un 45% y en 3,7% por la variable CF y IED, respectivamente. Observamos hasta el último dato generado en la tabla, el trimestre diez, nos muestra que la variable PIB es impulsada un 12,57% por la variable CF y un 1,54% por IED.

**Causalidad de Granger.**

**Hipótesis nula:** Las variables independientes CF y IED, no causan en el sentido de Granger a la variable dependiente PIB, durante el periodo estudiado:

**Gráfico 24. Causalidad de Granger**

**CF:** Al obtener una probabilidad menor al 5%, rechazamos la hipótesis nula, por lo que podemos afirmar que la variable independiente CF, sí causa en el sentido de Granger a la variable dependiente PIB.

**IED:** Se obtiene una probabilidad mayor al 5%, por lo que no se rechaza la hipótesis nula y se puede concluir que, basados en el presente modelo, la variable independiente IED, no causa en el sentido de Granger a la variable dependiente PIB.

**ALL:** El valor p para ambas constantes (ALL), es menor al 5%, por lo que se rechaza la hipótesis nula, podemos establecer que las variables independientes CF y IED, sí causan en el sentido de Granger a la variable dependiente PIB, durante el periodo estudiado.

## Discusión

### Contraste de los resultados con la teoría económica

La teoría de la Internalización (OLI) plantea que la inversión extranjera directa (IED) tiene un efecto positivo en el crecimiento económico de las naciones tanto en el corto como en el largo plazo. Los resultados del análisis econométrico realizado en el contexto de Chile respaldan esta afirmación al demostrar que tanto la Inversión Nacional (CF) como la Inversión Extranjera Directa (IED) tienen un impacto significativo en el Producto Interno Bruto (PIB) del país. Sin embargo, el efecto de la IED en el crecimiento económico no se muestra de manera sostenida en el tiempo, a diferencia de lo sugerido por la teoría OLI, ya que el modelo de impulso-respuesta no muestra una respuesta significativa del PIB a los impulsos de la IED. Además, los estudios previos en otros países mencionan la importancia del capital humano en la absorción de conocimientos tecnológicos transmitidos por empresas extranjeras, y aunque no se analiza directamente en el modelo, se podría considerar como una posible explicación de la falta de sostenibilidad de la relación entre IED y PIB en Chile.

Por otro lado, el rol de la acumulación de capital privado nacional, señalado por Romero (2012), no fue analizado en este estudio específico, pero es un aspecto relevante para considerar en futuras investigaciones. En comparación con el caso de Perú, donde la IED ha demostrado ser un motor del crecimiento económico en sectores estratégicos, en Chile se observa una relación significativa entre la Inversión Nacional y el PIB, lo que sugiere que la acumulación de capital privado nacional podría ser más relevante en el fomento del crecimiento económico del país.

## Conclusión

En conclusión, aunque la teoría OLI sostiene que la IED tiene un impacto positivo en el crecimiento económico, los resultados del análisis específico para Chile sugieren que otros factores, como la absorción de conocimientos tecnológicos y la acumulación de capital privado nacional, también desempeñan un papel relevante en el crecimiento económico del país.

## RESULTADOS

El modelo tiene un R-squared de 69%, lo que indica que aproximadamente el 69% de la variabilidad del PIB es explicada por las variables independientes. El valor del estadístico F de 0.00000 confirma la significancia conjunta del modelo y sus variables.

El análisis econométrico sobre la incidencia de la Inversión Extranjera Directa (IED) y la Inversión Nacional (CF) en el crecimiento económico de Chile revela que ambas variables tienen un impacto significativo en el Producto Interno Bruto (PIB). La constante (C) se mantiene negativa y no es representativa en el modelo. Por cada unidad de dólar que aumenta la Inversión Nacional, el PIB incrementa en un 0.08190, mientras que por cada unidad de dólar invertido en Inversión Extranjera Directa, el PIB aumenta en 0.0120.

La función de impulso-respuesta muestra que el PIB responde positivamente a los impulsos de la Inversión Nacional en los dos primeros periodos, pero este efecto no se mantiene en el tiempo. Por otro lado, la Inversión Extranjera Directa no presenta una respuesta significativa en el PIB.

La prueba de causalidad de Granger muestra que la Inversión Nacional causa en el sentido de Granger al PIB, mientras que la Inversión Extranjera Directa no presenta una relación causal significativa durante el periodo estudiado, el valor p para ambas variables, es 0.028, menor al 5%, por lo que se rechaza la hipótesis nula, podemos establecer que el PIB, es causado por la IN y la IED, basados en la causalidad de Granger.

# CONCLUSIÓN

El presente trabajo de investigación ha plateado como objetivo principal determinar la incidencia de la inversión extranjera directa y la inversión nacional en el crecimiento económico de Chile, acompañado de tres objetivos específicos como componentes. Dentro del apartado de resultado obtenidos por los procesos econométricos se verifica que las variables económicas están correlacionadas de forma positiva cumpliendo con la intención del proyecto, de esta manera que respalda a lo establecido en los objetivos.

Mediante al análisis de las variables independientes y las variables dependientes, la aplicación de funciones de transferencia, el método de Fisher sobre modelos dinámicos y los vectores autorregresivos, se cumplió con lo mencionado dentro de los objetivos específicos. Se constata que el crecimiento económico (PIB) de Chile, está influenciado por las variables de inversión nacional e inversión extranjera directa, de este modo se manifiesta que ambas inversiones pueden cusas efectos tanto positivos como negativos en la variación del PIB.

Sin embargo, el análisis econométrico de los procesos también resalta que el efecto de la Inversión Nacional en el PIB no es sostenido en el tiempo, y la IED no muestra una respuesta significativa en el PIB a lo largo del 2008 hasta 2023. Estos resultados resaltan la importancia de ambos tipos de inversión en el crecimiento económico de Chile, lo que resultaría como resultaría como el efecto de ganar y ganar, creando mayor dinamismo para la economía chilena y más oportunidades de trabajo y progreso para la sociedad.

En conclusión, para impulsar el crecimiento económico sostenible en Chile, es necesario seguir atrayendo inversiones extranjeras y fomentar la inversión interna. Además, se deben implementar políticas que promuevan la transferencia de tecnología y el desarrollo del capital humano para maximizar los beneficios de la inversión extranjera directa. El análisis econométrico realizado en este estudio proporciona una base sólida para la toma de decisiones y el diseño de políticas que impulsen el desarrollo económico del país.

# REFERENCIAS BIBLIOGRÁFICAS

Armijos, J., & Olaya, E. (2017). Efecto de la inversión extranjera directa en el crecimiento económico en Ecuador durante 1980-2015: un análisis de cointegración. 2, 1, 31-38. Revista Económica. https://revistas.unl.edu.ec/index.php/economica/article/view/205

Bustamante, F. (2017). La inversión extranjera directa en el Perú y sus implicancias en el crecimiento económico 2009-2015. 21, 2, 51-63. Peru: Pensamiento Crítico. https://doi.org/https://doi.org/10.15381/pc.v21i2.13259

CEPAL. (2021). Estudio Económico de América Latina y el Caribe. 9. https://repositorio.cepal.org/bitstream/handle/11362/47192/64/EE2021_Chile_es.pdf

Dunning, J. (s.f.). La producción internacional y la empresa multinacional. Londres: Allen & Urwin. https://dialnet.unirioja.es/descarga/articulo/4780130.pdf

Loja Barbecho, L. C., & Torres Guzmán, O. N. (Octubre de 2013). La inversión extranjera directa en el Ecuador durante el periodo 1979-2011: análisis de su incidencia en el crecimiento economico. 177. Cuenca, Ecuador. http://dspace.ucuenca.edu.ec/handle/123456789/4728

Romero, J. (2012). Inversión extranjera directa y crecimiento económico en México: 1940-2011. 71, 282, 109-147. Retrieved 24 de 07 de 2023, from http://www.scielo.org.mx/scielo.php?script=sci_arttext&pid=S0185-16672012000400005&lng=es&tlng=es.

## ANEXOS

*Anexo 1.*

| Periodo | 1.PIB a precios corrientes (miles de millones $) | 2.Formación bruta capital fijo (miles de millones $ encadenado) | 3.Inversión extranjera Directa |
|---|---|---|---|
| mar.2008 | 24.088,41 | 8.270,65 | 8.949,62 |
| jun.2008 | 23.898,72 | 8.742,26 | 9.685,98 |
| sep.2008 | 21.977,66 | 8.684,60 | 10.475,22 |
| dic.2008 | 23.902,33 | 9.346,91 | 11.172,72 |
| mar.2009 | 23.247,62 | 7.902,98 | 11.298,35 |
| jun.2009 | 23.481,16 | 7.554,55 | 11.036,33 |
| sep.2009 | 22.980,23 | 7.345,03 | 10.734,01 |
| dic.2009 | 26.429,46 | 8.366,79 | 12.326,37 |
| mar.2010 | 25.685,17 | 7.947,05 | 13.124,68 |
| jun.2010 | 26.801,71 | 8.320,24 | 14.604,19 |
| sep.2010 | 27.399,25 | 8.436,93 | 15.486,11 |
| dic.2010 | 30.891,74 | 9.927,02 | 15.697,19 |
| mar.2011 | 29.932,27 | 9.491,81 | 16.380,98 |
| jun.2011 | 29.967,72 | 9.553,24 | 17.076,29 |
| sep.2011 | 28.904,55 | 9.522,10 | 17.541,21 |
| dic.2011 | 32.704,77 | 11.281,61 | 18.751,25 |
| mar.2012 | 31.759,25 | 10.177,04 | 19.601,40 |
| jun.2012 | 32.084,08 | 10.422,80 | 20.774,14 |
| sep.2012 | 31.078,56 | 11.326,72 | 24.162,24 |
| dic.2012 | 35.051,50 | 12.708,04 | 29.870,07 |
| mar.2013 | 33.441,83 | 10.967,24 | 30.704,85 |
| jun.2013 | 33.908,57 | 11.318,71 | 32.585,05 |
| sep.2013 | 32.986,38 | 11.291,45 | 35.944,94 |
| dic.2013 | 36.972,40 | 11.647,09 | 38.388,32 |
| mar.2014 | 36.033,28 | 10.790,01 | 39.583,97 |
| jun.2014 | 36.551,42 | 10.731,37 | 41.399,88 |
| sep.2014 | 35.395,96 | 10.240,17 | 42.311,45 |
| dic.2014 | 39.970,64 | 11.629,60 | 47.132,36 |
| mar.2015 | 39.274,10 | 10.440,25 | 47.297,40 |
| jun.2015 | 39.179,28 | 10.378,67 | 48.344,20 |
| sep.2015 | 37.988,16 | 10.944,55 | 55.605,61 |
| dic.2015 | 42.181,37 | 11.674,08 | 55.958,29 |
| mar.2016 | 42.065,86 | 10.735,08 | 57.310,84 |
| jun.2016 | 41.209,40 | 10.300,10 | 57.556,91 |
| sep.2016 | 40.601,42 | 10.184,48 | 58.366,83 |
| dic.2016 | 44.888,01 | 11.192,78 | 57.715,17 |
| mar.2017 | 43.544,74 | 10.038,59 | 56.361,08 |
| jun.2017 | 43.842,77 | 9.769,70 | 57.901,82 |
| sep.2017 | 43.571,82 | 9.811,49 | 57.861,29 |
| dic.2017 | 48.355,58 | 11.382,32 | 58.482,31 |
| mar.2018 | 46.966,37 | 10.701,49 | 60.057,95 |
| jun.2018 | 46.991,41 | 10.547,39 | 54.884,68 |
| sep.2018 | 45.059,32 | 10.412,86 | 55.275,59 |
| dic.2018 | 50.417,77 | 12.002,61 | 56.859,48 |
| mar.2019 | 48.582,23 | 11.146,25 | 56.450,26 |
| jun.2019 | 48.552,83 | 11.046,02 | 57.179,03 |
| sep.2019 | 47.469,51 | 11.012,96 | 56.886,76 |
| dic.2019 | 51.147,66 | 12.417,69 | 57.388,73 |
| mar.2020 | 51.350,09 | 11.307,58 | 56.414,84 |
| jun.2020 | 46.078,92 | 8.939,44 | 57.336,01 |
| sep.2020 | 47.327,27 | 9.230,57 | 57.534,06 |
| dic.2020 | 56.672,62 | 11.202,67 | 57.401,57 |
| mar.2021 | 56.251,52 | 11.223,92 | 56.550,10 |
| jun.2021 | 57.580,57 | 10.693,42 | 55.861,08 |
| sep.2021 | 59.343,87 | 11.903,69 | 56.175,61 |
| dic.2021 | 67.195,51 | 13.250,77 | 55.630,23 |
| mar.2022 | 64.083,30 | 12.047,15 | 57.446,00 |
| jun.2022 | 63.745,50 | 11.356,55 | 58.079,03 |
| sep.2022 | 63.841,27 | 11.959,95 | 58.294,13 |
| dic.2022 | 70.923,29 | 13.031,58 | 56.871,92 |
| mar.2023 | 70.125,93 | 11.789,44 | 53.263,98 |

*Anexo 2. Ecuacion de regresión lineal*

**Gráfico 1. Ecuación de regresión lineal**

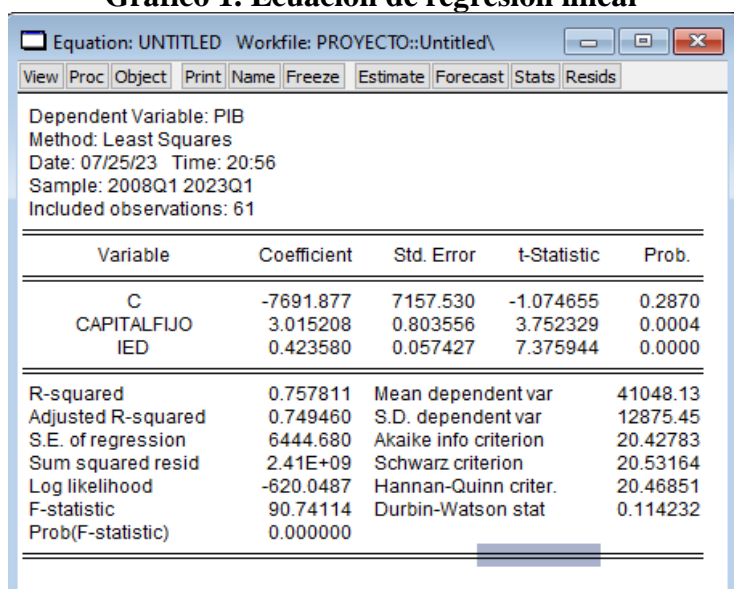

**Fuente: Los Autores, 2023**

*Anexo 3. Regresión Lineal Con Series De Tiempo, Función De Transferencia*

Para realizar un ajuste del modelo válido en el largo plazo será necesario que exista una relación de cointegración entre las variables del modelo. Para ello:

En primer lugar, analizaremos la estacionariedad de todas las variables que lo integran. En primer lugar, representamos las series mediante Quick → Graph → Line Graph y rellenando la pantalla Series List sucesivamente con las variables PIB, IED y CAPITAL FIJO. Se obtienen los siguientes gráficos:

**Gráfico 25. Gráfico de la variable PIB**

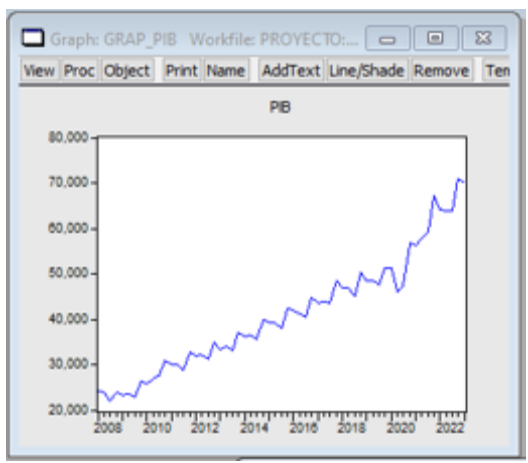

*Fuente: Los Autores, 2023.*

**Gráfico 25. Gráfico de la variable CAPITAL FIJO**

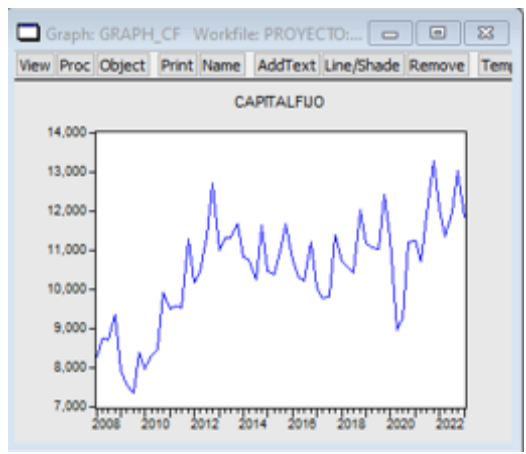

*Fuente: Los Autores, 2023.*

**Gráfico 26. Gráfico de la variable CAPITAL FIJO**

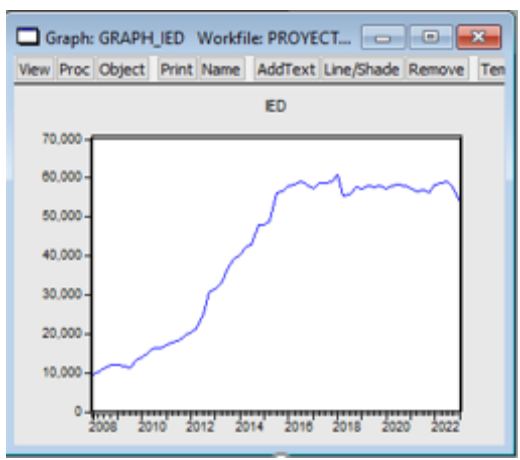

*Fuente: Los Autores, 2023.*

- A simple vista se observa la ausencia de estacionariedad de las variables. Pero para poder afirmar este hecho, es necesario utilizar contrastes formales, por el ejemplo el test de raíces unitarias de Phillips Perron, Para llevar a cabo este contraste desde Eviews elegimos Quick Series Statistics Unit Roots Tests, rellenamos la pantalla Series Name con la variable 'Y' y al pulsar OK se obtiene la pantalla Unit Root Tests en cuyo campo Test Type elegimos Phillips-Perron y en cuyo campo Test for unit root in elegimos Level que estamos probando la estacionariedad de la serie Yen niveles.

**Gráfico 27. Paso de raíces unitaria Phillips Perron**

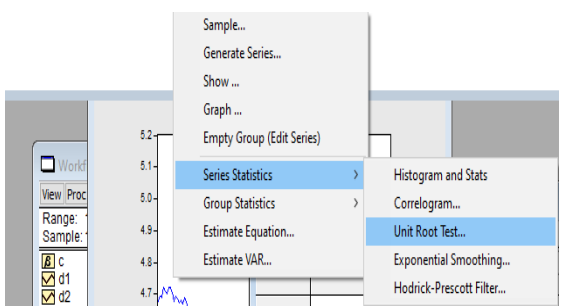

*Fuente: Los Autores, 2023.*

**Gráfico 28. Paso de raíces unitaria Phillips Perron**

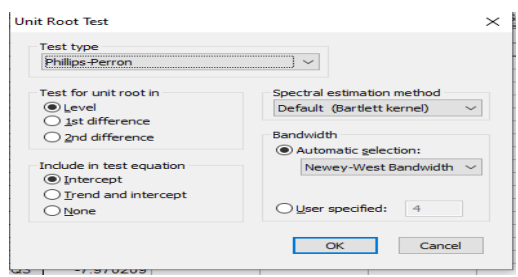

*Fuente: Los Autores, 2023.*

**Gráfico 2. Test de Phillips-Perron – PIB**

*Fuente: Los Autores, 2023.*

**Gráfico 3. Test de Phillips-Perron – CAPITAL FIJO**

*Fuente: Los Autores, 2023.*

**Gráfico 4. Test de Phillips-Perron – IED**

*Fuente: Los Autores, 2023.*

Al pulsar OK se obtiene un p valor mayor que 0.05 en los resultados de las variables PIB, IED, CAPITAL FIJO con el contraste de Phillips Perron, lo que indica ausencia de estacionariedad. Lo que nos lleva a considerar sus primeras diferencias.

- A continuación, estudiaremos la estacionariedad de las primeras diferencias de las variables del modelo mediante el contraste de Phillips Perron. Elegimos Quick → Series Statistics Unit Root Test, rellenamos la pantalla Series Name con la variable 'Y' y al pulsar OK se obtiene la pantalla Unit Root Test en cuyo campo Test Type elegimos Phillips-Perron y en cuyo campo Test for unit root in elegimos 1st Difference ya que estamos probando la estacionariedad de la primera diferencia de la serie.

**Gráfico 29. Paso de raíces unitaria Phillips Perron (Primera diferencia)**

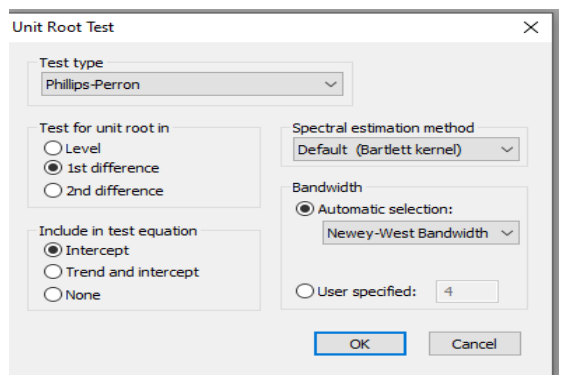

*Fuente: Los Autores, 2023.*

**Gráfico 5. Test de Phillips-Perron - PIB (estacionariedad corregida)**

| Series: PIB   Workfile: PROYECTO::Untitled\ |
| --- |

View Proc Object Properties  Print Name Freeze  Sample Genr Sheet Graph Stats I

**Phillips-Perron Unit Root Test on D(PIB)**

Null Hypothesis: D(PIB) has a unit root
Exogenous: Constant
Bandwidth: 28 (Newey-West using Bartlett kernel)

|  |  | Adj. t-Stat | Prob.* |
| --- | --- | --- | --- |
| Phillips-Perron test statistic |  | -12.03542 | 0.0000 |
| Test critical values: | 1% level | -3.546099 |  |
|  | 5% level | -2.911730 |  |
|  | 10% level | -2.593551 |  |

*MacKinnon (1996) one-sided p-values.

| Residual variance (no correction) |  | 6600826. |
| --- | --- | --- |
| HAC corrected variance (Bartlett kernel) |  | 3647100. |

*Fuente: Los Autores, 2023.*

**Gráfico 6. Test de Phillips-Perron – CAPITAL FIJO (estacionariedad corregida)**

*Fuente: Los Autores, 2023.*

**Gráfico 7. Test de Phillips-Perron – IED (estacionariedad corregida)**

*Fuente: Los Autores, 2023.*

Se obtiene un p-valor menor que 0,05 en los resultados del contraste de Phillips Perron, lo que indica estacionariedad en la primera diferencia de las variables Y, X2, X3.

- Hemos visto que las tres variables del modelo son no estacionarias, pero sí lo son sus primeras diferencias. Esto indica que las tres series son I (1). Por lo tanto, se cumple la primera condición para que exista una relación de cointegración.

- La siguiente tarea será comprobar que efectivamente las variables cointegran. Para ello ajustamos el modelo Y = A + BX2+ BX3 y comprobamos si los residuos estimados tienen raíces unitarias (son estacionarios). Para ello usamos Quick Estimate Equation, escribiendo la ecuación del modelo a ajustar en el campo. Equation Specification de la solapa Specification, eligiendo Least Squares en el campo Method para ajustar para mínimos cuadrados y haciendo clic en Aceptar.

**Gráfico 29. Estimación de la ecuación**

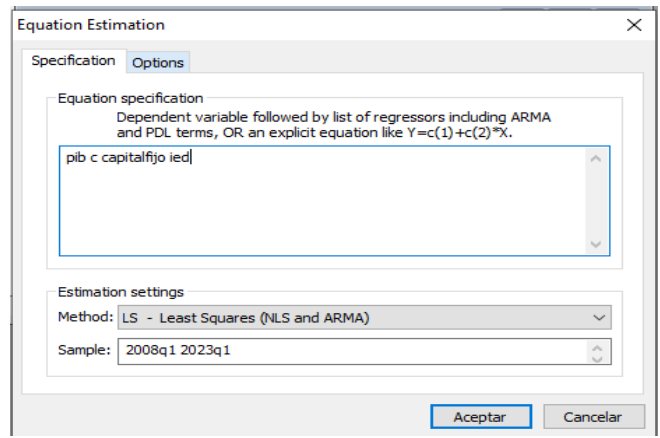

*Fuente: Los Autores, 2023.*

**Gráfico 1. Ecuación de regresión lineal**

Dependent Variable: PIB
Method: Least Squares
Date: 07/25/23   Time: 20:56
Sample: 2008Q1 2023Q1
Included observations: 61

| Variable | Coefficient | Std. Error | t-Statistic | Prob. |
|---|---|---|---|---|
| C | -7691.877 | 7157.530 | -1.074655 | 0.2870 |
| CAPITALFIJO | 3.015208 | 0.803556 | 3.752329 | 0.0004 |
| IED | 0.423580 | 0.057427 | 7.375944 | 0.0000 |
| R-squared | 0.757811 | Mean dependent var | | 41048.13 |
| Adjusted R-squared | 0.749460 | S.D. dependent var | | 12875.45 |
| S.E. of regression | 6444.680 | Akaike info criterion | | 20.42783 |
| Sum squared resid | 2.41E+09 | Schwarz criterion | | 20.53164 |
| Log likelihood | -620.0487 | Hannan-Quinn criter. | | 20.46851 |
| F-statistic | 90.74114 | Durbin-Watson stat | | 0.114232 |
| Prob(F-statistic) | 0.000000 | | | |

*Fuente: Los Autores, 2023.*

Se obtiene el modelo ajustado de la Figura. Para guardar los residuos estimados (variable RESID) con otro nombre, seleccionamos la variable RESID, elegimos Object Copy Selected y en la pantalla Object Copy elegimos RESID01 como variable destino.

**Gráfico 30. Copia del Resid**

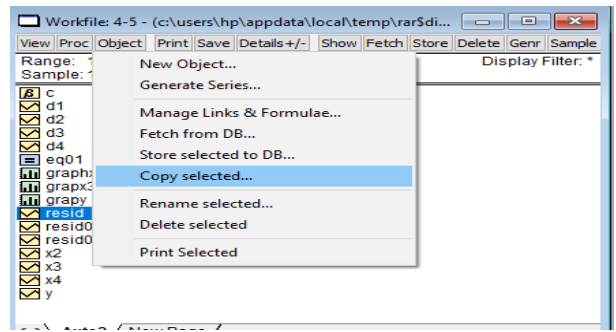

*Fuente: Los Autores, 2023.*

**Gráfico 30. Copia del Resid**

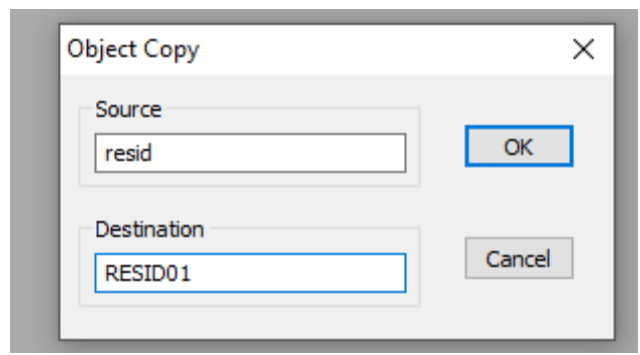

*Fuente: Los Autores, 2023.*

Al pulsar OK ya tenemos RESIDO1 como nueva variable copia de RESID.

- Para comprobar que los residuos estimados del ajuste anterior son estacionarios elegimos Quick Series Statistics Unit Root Test, rellenamos la pantalla Series Name con la variable RESIDO1 y al pulsar OK se obtiene la pantalla Unit Root Tests en cuyo campo Test Type elegimos Phillips-Perron y en cuyo campo Test for unit root in elegimos Level ya que estamos probando la estacionariedad de la serie RESID01 en niveles.

**Gráfico 31. Test de raíces unitarias en Resid01**

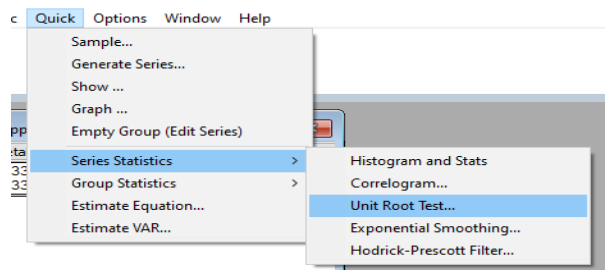

*Fuente: Los Autores, 2023.*

**Gráfico 32. Test de raíces unitarias en Resid01(a nivel)**

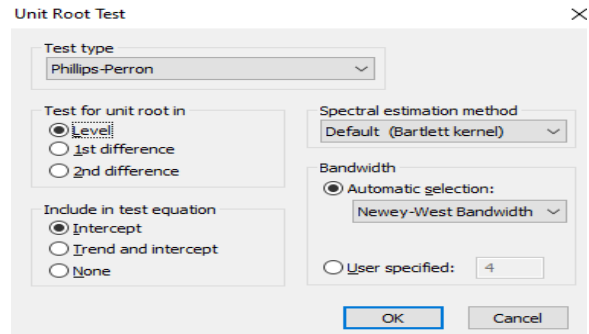

*Fuente: Los Autores, 2023.*

**Gráfico 8. Test de Phillips-Perron – Resid01**

Series: RESID01  Workfile: PROYECTO::Untitled\

View Proc Object Properties | Print Name Freeze | Sample Genr Sheet Graph Stats |

**Phillips-Perron Unit Root Test on RESID01**

Null Hypothesis: RESID01 has a unit root
Exogenous: Constant
Bandwidth: 4 (Newey-West using Bartlett kernel)

|  |  | Adj. t-Stat | Prob.* |
|---|---|---|---|
| Phillips-Perron test statistic |  | 0.866459 | 0.9944 |
| Test critical values: | 1% level | -3.544063 |  |
|  | 5% level | -2.910860 |  |
|  | 10% level | -2.593090 |  |

*MacKinnon (1996) one-sided p-values.

| Residual variance (no correction) | 4481992. |
|---|---|
| HAC corrected variance (Bartlett kernel) | 3828350. |

Phillips-Perron Test Equation
Dependent Variable: D(RESID01)
Method: Least Squares

*Fuente: Los Autores, 2023.*

Al pulsar OK se obtiene un p-valor mayor que 0.05 en los resultados del contraste de Phillips Perron, lo que indica ausencia de estacionariedad en los residuos del modelo. Llegamos entonces a la conclusión de que las variables del modelo no cointegran y este puede ser espurio.

- Una vez vista la no existencia de una relación de cointegración, hemos llegado a la conclusión de que el ajuste Y=B+ BX2+6X3+u, cuyos resultados se presentan puede ser espurio.

  Aunque estos resultados presentan una gran significatividad individual y conjunta de los parámetros estimados del modelo y un alto coeficiente de determinación ajustado, se observa que el bajo valor del estadístico de Durban Watson denota autocorrelación serial. Este hecho, junto con la ausencia de cointegración pueden llevarnos a pensar en la presencia de cambios estructurales.

Para detectar el posible cambio estructural, sobre los resultados de la regresión elegimos View Stability Tests Recursive Estimates (OLS only) según la y en la pantalla Recursive Estimation elegimos Recursive Residuals.

**Gráfico 33. Tests Recursive Estimates**

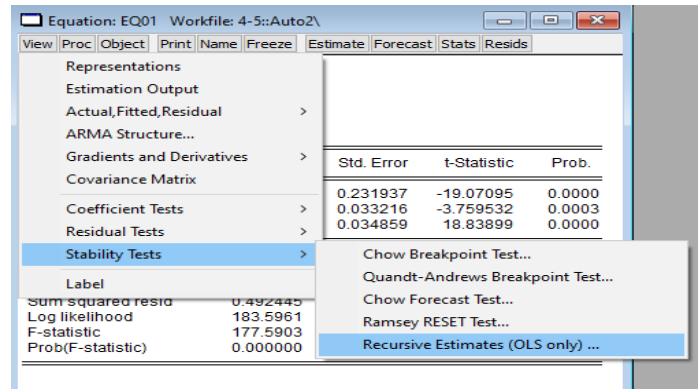

*Fuente: Los Autores, 2023.*

**Gráfico 34. Tests Recursive Estimates**

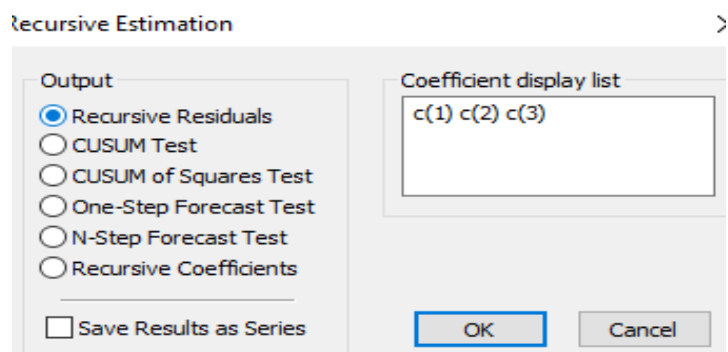

*Fuente: Los Autores, 2023.*

**Gráfico 35. Grafica Recursive Residual**

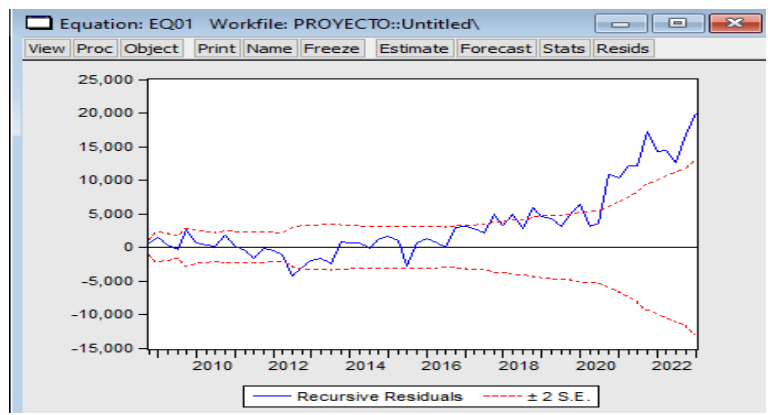

*Fuente: Los Autores, 2023.*

Al pulsar OK se obtiene el gráfico de residuos recursivos que detecta posibles cambios estructurales alrededor de 2012, 2017, 2021 y 2022.

- Ahora introduciremos una variable dicotómica ´´d1´´ por cada cambio estructural, que valga cero antes de la fecha del cambio y que valga uno después de esa fecha Como 2021 y 2022 están muy cerca, consideraremos solamente los cambios de 2012, 2017 y 2021.

$$D_i = \begin{cases} 0 & si \quad i < t \\ 1 & si \quad i \geq t \end{cases}$$

Ajustamos el nuevo modelo por MCO y guardamos los nuevos residuos como ''RESID02''

**Gráfico 36. Estimación de la ecuación con las variables dicotómicas.**

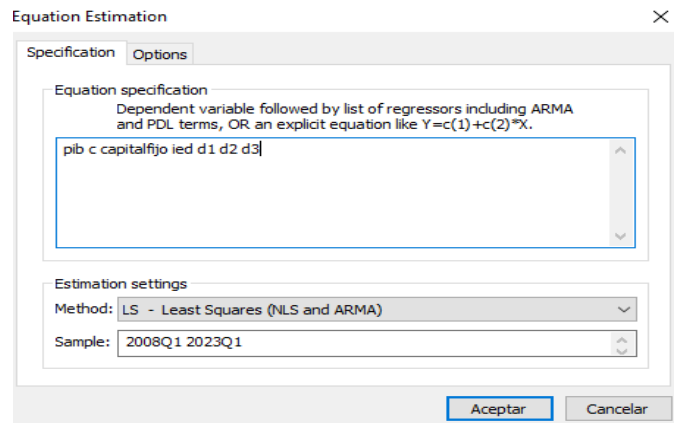

*Fuente: Los Autores, 2023.*

**Gráfico 37. Regresión con las variables dicotómicas.**

Dependent Variable: PIB
Method: Least Squares
Date: 07/25/23   Time: 21:12
Sample: 2008Q1 2023Q1
Included observations: 61

| Variable | Coefficient | Std. Error | t-Statistic | Prob. |
|---|---|---|---|---|
| C | 3442.796 | 3440.990 | 1.000525 | 0.3214 |
| CAPITALFIJO | 2.151688 | 0.381086 | 5.646204 | 0.0000 |
| IED | 0.300825 | 0.042933 | 7.006800 | 0.0000 |
| D1 | -2527.361 | 1681.488 | -1.503050 | 0.1385 |
| D2 | 6773.382 | 1061.770 | 6.379332 | 0.0000 |
| D3 | 13359.80 | 1135.905 | 11.76138 | 0.0000 |
| R-squared | 0.965973 | Mean dependent var | | 41048.13 |
| Adjusted R-squared | 0.962880 | S.D. dependent var | | 12875.45 |
| S.E. of regression | 2480.651 | Akaike info criterion | | 18.56361 |
| Sum squared resid | 3.38E+08 | Schwarz criterion | | 18.77124 |
| Log likelihood | -560.1901 | Hannan-Quinn criter. | | 18.64498 |
| F-statistic | 312.2769 | Durbin-Watson stat | | 1.274571 |
| Prob(F-statistic) | 0.000000 | | | |

*Fuente: Los Autores, 2023.*

Para comprobar si los residuos del nuevo ajuste son estacionarios elegimos el test de Philips-Perron.

**Gráfico 38. Regresión con las variables dicotómicas.**

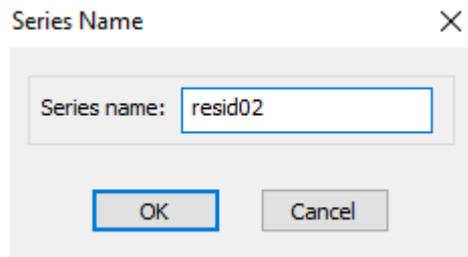

*Fuente: Los Autores, 2023.*

**Gráfico 9. Test de Phillips-Perron – Resid02 (Corregido)**

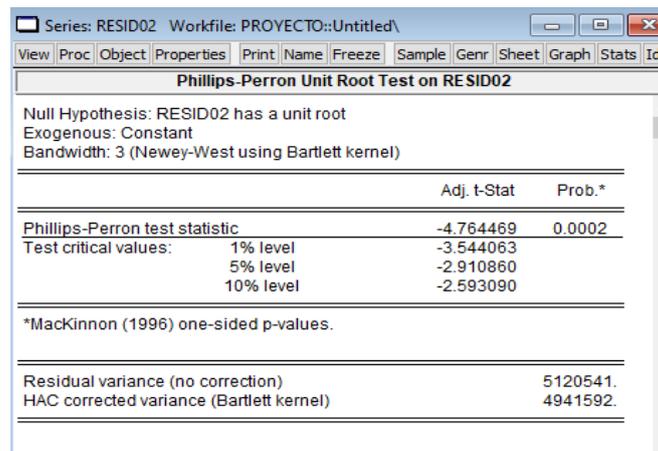

*Fuente: Los Autores, 2023.*

Se obtiene un valor menor que 0.05 en el resultado de Phillips-Perron. Lo que nos indica presencia de estacionariedad en los residuos del modelo. Llegamos entonces a la conclusión de que las variables del nuevo modelo cointegran y ya no es espurio.

La cointegración estimada será:

$Y = 3442.796 + 2.151688\,CF + 0.300825\,IED - 2527.361\,D1 + 6773.382\,D2 + 13359.80\,D3$

## 1.- Heterocedasticidad

Procedemos a comprobar si el modelo tiene problemas de heterocedasticidad mediante el test de Glejser.

### Prueba de Glejser

Se escoge el supuesto para verificar la ausencia o presencia de heterocedasticidad en nuestro modelo mediante el test de Glejser.

**Gráfico 10. Test de Glejser**

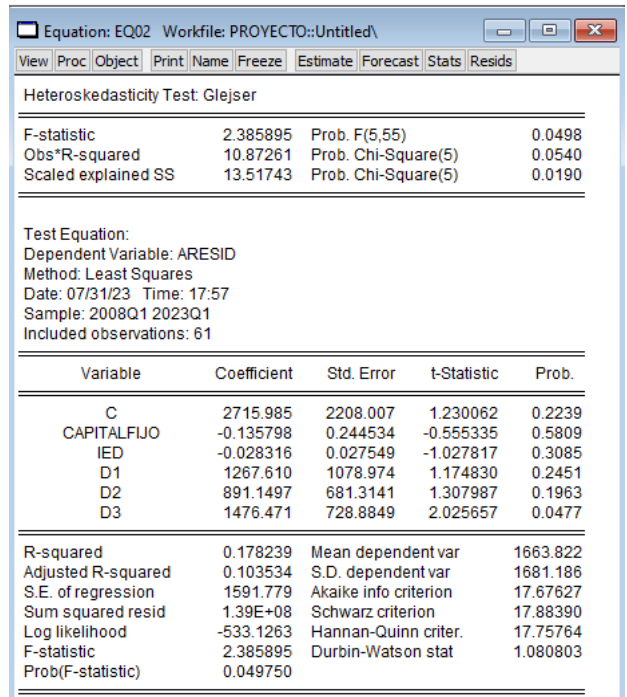

*Fuente: Los Autores, 2023.*

La probabilidad de F es menor al 5%, por ende, se confirma que el modelo no tiene problemas de heterocedasticidad, por ende, no es necesaria su corrección

### 2.- Normalidad

Se verifica la ausencia o presencia de normalidad en nuestro modelo mediante el análisis de Jarque-Bera.

**Gráfico 11. Test de Jarque-Bera**

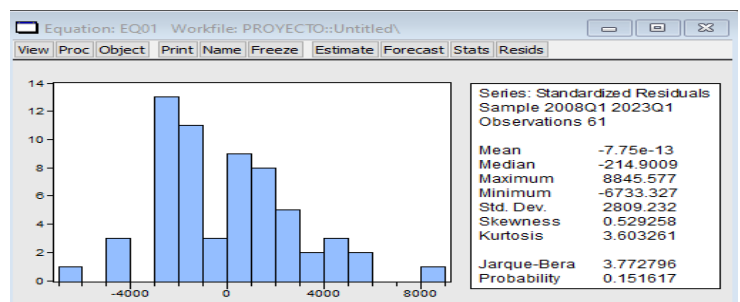

*Fuente: Los Autores, 2023.*

El resultado de Jarque-Bera en la tabla es 0.1516 En el caso de nuestro modelo, la probabilidad de Jarque-Bera es mayor al 5%, por lo tanto, nuestro modelo tiene normalidad en los residuos.

### 3.- Linealidad

Se verifica la ausencia o presencia de linealidad en nuestro modelo mediante el análisis de Ramsey RESET.

**Gráfico 12. Test de Ramsey Reset**

**Ramsey RESET Test:**

| | | | |
|---|---|---|---|
| F-statistic | 2.898430 | Prob. F(1,54) | 0.0944 |
| Log likelihood ratio | 3.189306 | Prob. Chi-Square(1) | 0.0741 |

Test Equation:
Dependent Variable: PIB
Method: Least Squares
Date: 07/31/23   Time: 18:00
Sample: 2008Q1 2023Q1
Included observations: 61

| Variable | Coefficient | Std. Error | t-Statistic | Prob. |
|---|---|---|---|---|
| C | 14674.13 | 7413.935 | 1.979263 | 0.0529 |
| CAPITALFIJO | 0.368579 | 1.112361 | 0.331348 | 0.7417 |
| IED | 0.075171 | 0.139103 | 0.540395 | 0.5911 |
| D1 | 484.1352 | 2421.164 | 0.199960 | 0.8423 |
| D2 | 136.2770 | 4035.845 | 0.033765 | 0.9732 |
| D3 | -3280.912 | 9838.005 | -0.333494 | 0.7401 |
| FITTED^2 | 1.06E-05 | 6.23E-06 | 1.702478 | 0.0944 |

| | | | |
|---|---|---|---|
| R-squared | 0.967707 | Mean dependent var | 41048.13 |
| Adjusted R-squared | 0.964119 | S.D. dependent var | 12875.45 |
| S.E. of regression | 2438.916 | Akaike info criterion | 18.54411 |
| Sum squared resid | 3.21E+08 | Schwarz criterion | 18.78635 |
| Log likelihood | -558.5955 | Hannan-Quinn criter. | 18.63905 |
| F-statistic | 269.6962 | Durbin-Watson stat | 1.204022 |
| Prob(F-statistic) | 0.000000 | | |

*Fuente: Los Autores, 2023.*

Si la probabilidad es menor al 5% existe problema de linealidad y al ser mayor al 5%, no existe.

### 4.- Autocorrelación

Se verifica la ausencia o presencia de autocorrelación en nuestro modelo mediante el análisis de estadístico Durbin-Watson.

**Gráfico 13. Estadístico Durbin-Watson**

Dependent Variable: PIB
Method: Least Squares
Date: 07/25/23   Time: 21:12
Sample: 2008Q1 2023Q1
Included observations: 61

| Variable | Coefficient | Std. Error | t-Statistic | Prob. |
|---|---|---|---|---|
| C | 3442.796 | 3440.990 | 1.000525 | 0.3214 |
| CAPITALFIJO | 2.151688 | 0.381086 | 5.646204 | 0.0000 |
| IED | 0.300825 | 0.042933 | 7.006800 | 0.0000 |
| D1 | -2527.361 | 1681.488 | -1.503050 | 0.1385 |
| D2 | 6773.382 | 1061.770 | 6.379332 | 0.0000 |
| D3 | 13359.80 | 1135.905 | 11.76138 | 0.0000 |

| | | | |
|---|---|---|---|
| R-squared | 0.965973 | Mean dependent var | 41048.13 |
| Adjusted R-squared | 0.962880 | S.D. dependent var | 12875.45 |
| S.E. of regression | 2480.651 | Akaike info criterion | 18.56361 |
| Sum squared resid | 3.38E+08 | Schwarz criterion | 18.77124 |
| Log likelihood | -560.1901 | Hannan-Quinn criter. | 18.64498 |
| F-statistic | 312.2769 | Durbin-Watson stat | 1.274571 |
| Prob(F-statistic) | 0.000000 | | |

*Fuente: Los Autores, 2023.*

Podemos visualizar que DW se encuentra en 1,27%. Por ende, podemos afirmar si hay problema de autocorrelación en nuestro modelo.

Se lo corrige aplicando un rezago en el modelo con la variable Ar (1)

**Gráfico 14. Corrección de Autocorrelación**

Dependent Variable: PIB
Method: Least Squares
Date: 07/31/23   Time: 18:03
Sample (adjusted): 2008Q2 2023Q1
Included observations: 60 after adjustments
Convergence achieved after 158 iterations

| Variable | Coefficient | Std. Error | t-Statistic | Prob. |
|---|---|---|---|---|
| C | 2768061. | 1.28E+08 | 0.021627 | 0.9828 |
| CAPITALFIJO | 2.494273 | 0.227443 | 10.96658 | 0.0000 |
| IED | -0.315447 | 0.126999 | -2.483854 | 0.0162 |
| D1 | 1220.924 | 1623.091 | 0.752221 | 0.4552 |
| D2 | 256.7456 | 1620.019 | 0.158483 | 0.8747 |
| D3 | -1589.958 | 1608.821 | -0.988275 | 0.3275 |
| AR(1) | 0.999688 | 0.014566 | 68.63214 | 0.0000 |
| R-squared | 0.986465 | Mean dependent var | | 41330.79 |
| Adjusted R-squared | 0.984932 | S.D. dependent var | | 12791.82 |
| S.E. of regression | 1570.195 | Akaike info criterion | | 17.66507 |
| Sum squared resid | 1.31E+08 | Schwarz criterion | | 17.90941 |
| Log likelihood | -522.9520 | Hannan-Quinn criter. | | 17.76064 |
| F-statistic | 643.7835 | Durbin-Watson stat | | 2.084228 |
| Prob(F-statistic) | 0.000000 | | | |

| Inverted AR Roots | 1.00 | | | |

*Fuente: Los Autores, 2023.*

Podemos visualizar que DW se encuentra en 2,84%. Por ende, podemos afirmar que se ha corregido el problema de autocorrelación en nuestro modelo.

### Anexo 4. Regresión Lineal Con Series De Tiempo, Modelos Dinámicos.

Se trata de ajustar, según el retardo aritmético de Fisher, el modelo siguiente:

$$Y_t = \mu + \sum_{i=0}^{7} \delta X_{t-i} + u_t$$

$$\text{Retardo aritmético de Fisher} \rightarrow \delta_i = \begin{cases} (7+1-i)\delta & 0 \leq i \leq 7 \\ 0 & i > 7 \end{cases}$$

$$Y_t = \mu + \sum_{i=0}^{7} \delta_i X_{t-i} + u_t = \mu + \delta \underbrace{\sum_{i=0}^{7} (7+1-i) X_{t-i}}_{ZF_t} + u_t = \mu + \delta ZF_t + u_t$$

Comenzamos generando la variable ZCF y ZIED. haciendo clic en GENR y rellenando la pantalla Generate Series by Equation según se indica. Al pulsar OK ya disponemos de las variables en nuestro fichero de trabajo. A continuación, ajustamos el modelo

$$Y_t = \mu + \delta ZF_t + u_t.$$

Para realizar el ajuste MCO con Eviews se elige Quick Estimate Equation, se escribe la ecuación del modelo a ajustar en el campo Equation Specification se elige Least Squares en el camp Method para ajustar por mínimos cuadrados y se hace clic en Aceptar.

**Gráfico 39. Generación de la variable ZCF.**

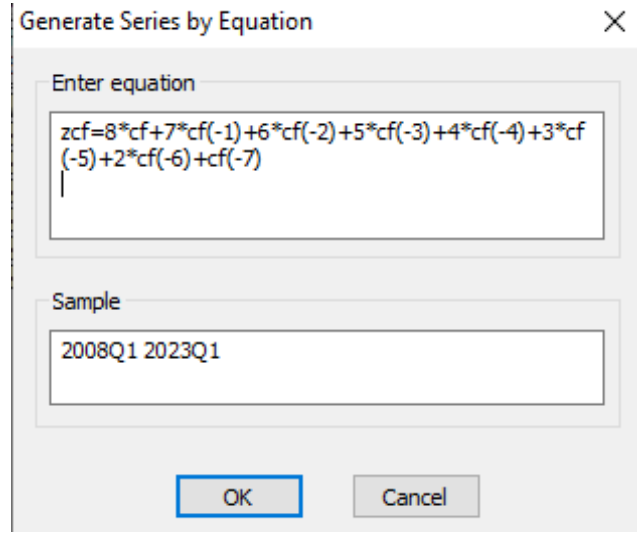

*Fuente: Los Autores, 2023.*

**Gráfico 40. Generación de la variable ZCF.**

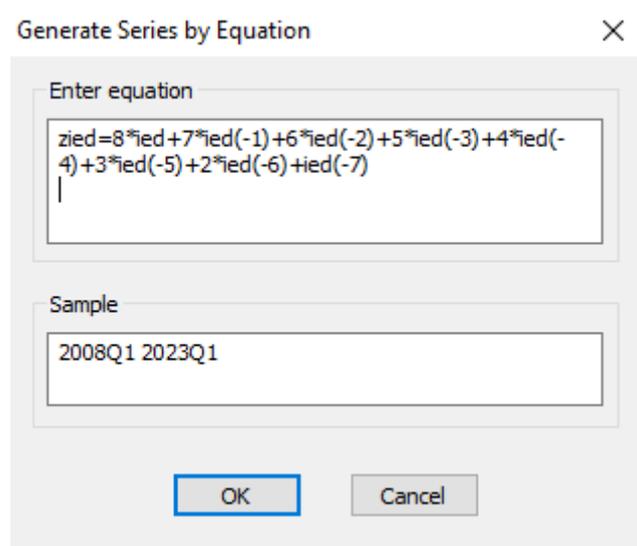

*Fuente: Los Autores, 2023.*

**Gráfico 41. Estimación de la ecuación ZCF.**

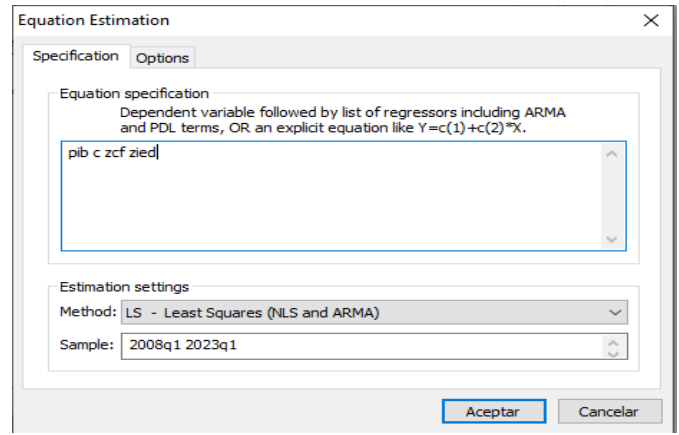

**Gráfico 15. Ecuación de regresión lineal (retardo dinámico de Fisher)**

Equation: UNTITLED   Workfile: PROYECTO2::Untitled\

View  Proc  Object  Print  Name  Freeze  Estimate  Forecast  Stats  Resids

Dependent Variable: PIB
Method: Least Squares
Date: 08/01/23   Time: 14:43
Sample (adjusted): 2009Q4 2023Q1
Included observations: 54 after adjustments

| Variable | Coefficient | Std. Error | t-Statistic | Prob. |
|----------|-------------|------------|-------------|-------|
| C | -6285.040 | 11386.60 | -0.551968 | 0.5834 |
| ZCF | 0.081906 | 0.034776 | 2.355255 | 0.0224 |
| ZIED | 0.012064 | 0.001990 | 6.063136 | 0.0000 |
| R-squared | 0.695788 | Mean dependent var | | 43340.00 |
| Adjusted R-squared | 0.683858 | S.D. dependent var | | 11873.30 |
| S.E. of regression | 6675.938 | Akaike info criterion | | 20.50436 |
| Sum squared resid | 2.27E+09 | Schwarz criterion | | 20.61486 |
| Log likelihood | -550.6177 | Hannan-Quinn criter. | | 20.54697 |
| F-statistic | 58.32319 | Durbin-Watson stat | | 0.156848 |
| Prob(F-statistic) | 0.000000 | | | |

El ajuste del modelo es bueno, salvo el problema de la autocorrelación residual derivada del valor tan bajo del estadístico de Durbin Watson. Ante esta situación intentaremos identificar la estructura de los residuos del modelo a través de View > Residual Tests > Correlogram Squared Residuals y tomando el número de retardos por defecto (54).

**Gráfico 42. Paso: Residual Test - Correlograma.**

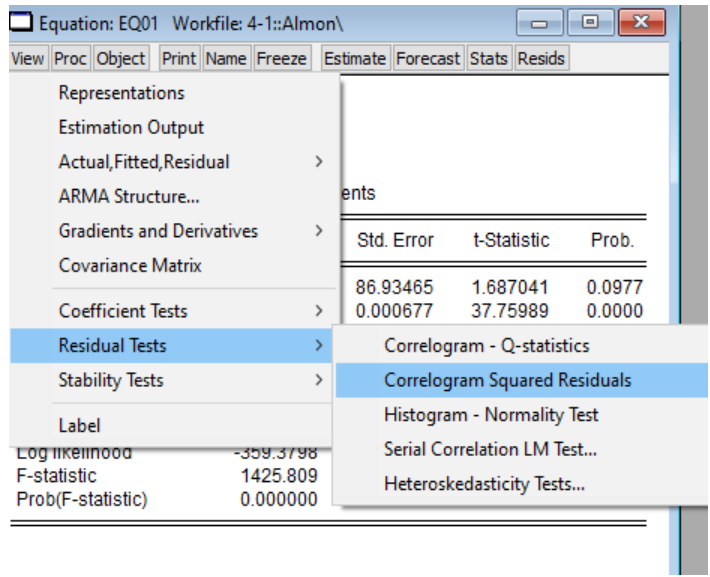

*Fuente: Los Autores, 2023.*

**Gráfico 43. Residual Test - Correlograma.**

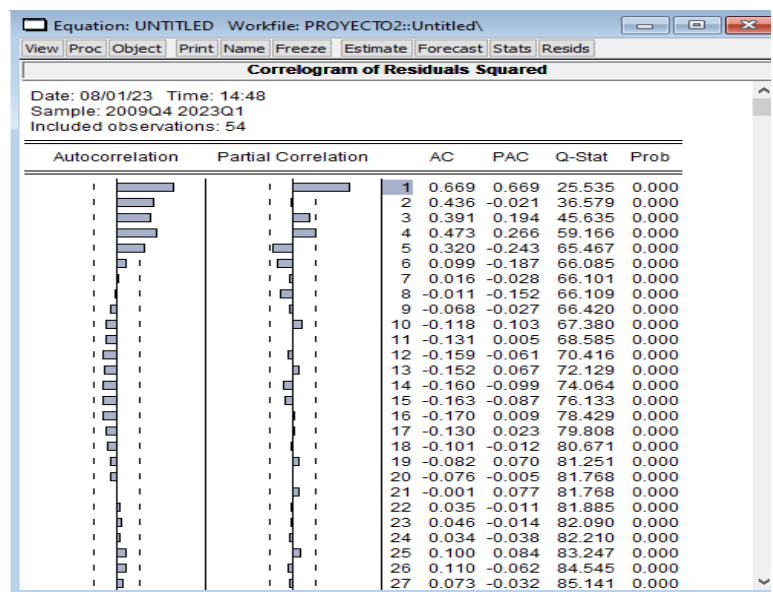

*Fuente: Los Autores, 2023.*

Se observa un decrecimiento de los términos de la FAC y los dos primeros retardos significativos en la FACP, lo que nos lleva a tomar una estructura autorregresiva de orden dos en los residuos.

$$Y_t = \mu + \delta Z F_t + u_t$$

Ajustaremos entonces el modelo con una estructura AR (2) en sus residuos. Se obtienen los resultados. El modelo presenta buena significatividad individual y conjunta de los parámetros estimados, altos coeficientes de determinación y un estadístico de Durbin Watson casi igual a 2, el ajuste es correcto

**Gráfico 44. Estimación de la ecuación con AR (1) y AR (2)**

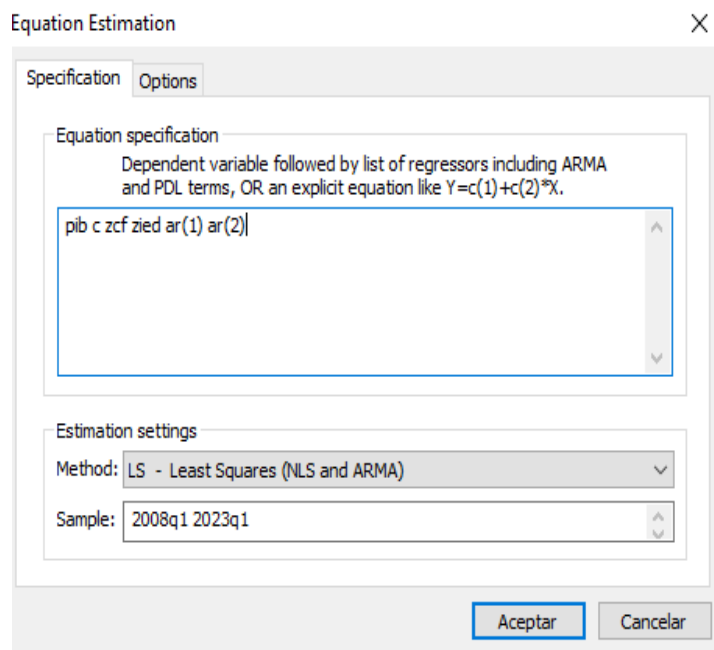

**Gráfico 17. Autocorrelación Corregida**

| Dependent Variable: PIB | | | | |
| --- | --- | --- | --- | --- |
| Method: Least Squares | | | | |
| Date: 08/01/23   Time: 14:53 | | | | |
| Sample (adjusted): 2010Q2 2023Q1 | | | | |
| Included observations: 52 after adjustments | | | | |
| Convergence achieved after 85 iterations | | | | |

| Variable | Coefficient | Std. Error | t-Statistic | Prob. |
| --- | --- | --- | --- | --- |
| C | -38253.68 | 119957.6 | -0.318893 | 0.7512 |
| ZCF | 0.137854 | 0.039814 | 3.462434 | 0.0012 |
| ZIED | -0.001831 | 0.013322 | -0.137439 | 0.8913 |
| AR(1) | 0.793639 | 0.145889 | 5.440025 | 0.0000 |
| AR(2) | 0.226586 | 0.155021 | 1.461653 | 0.1505 |
| R-squared | 0.956823 | Mean dependent var | | 44004.71 |
| Adjusted R-squared | 0.953148 | S.D. dependent var | | 11590.27 |
| S.E. of regression | 2508.753 | Akaike info criterion | | 18.58417 |
| Sum squared resid | 2.96E+08 | Schwarz criterion | | 18.77179 |
| Log likelihood | -478.1884 | Hannan-Quinn criter. | | 18.65610 |
| F-statistic | 260.3833 | Durbin-Watson stat | | 1.997213 |
| Prob(F-statistic) | 0.000000 | | | |

| Inverted AR Roots | 1.02 | | -.22 | |
| --- | --- | --- | --- | --- |
| | Estimated AR process is nonstationary | | | |

- Se observa que la estimación del parámetro S es 0.137854 para lo que es ZCF. Entonces tenemos:

$$\delta_i = \begin{cases} (7+1-i)\delta & 0 \le i \le 7 \\ 0 & i > 7 \end{cases} = \{8\delta, 7\delta, 6\delta, 5\delta, 4\delta, 3\delta, 2\delta, 1\delta\} =$$

$$\{1,102832;\ 0,964978;\ 0,827124;\ 0,68927;\ 0,551416;\ 0,413562;\ 0,27508;\ 0,137854\}$$

Por lo tanto, la estimación del modelo inicial será la siguiente:

$$Y_t = -38253,68 + 1,102832\ CF_t + 0,964978\ CF_{t-1} + 0,827124\ CF_{t-2}$$
$$+ 0,68927\ CF_{t-3} + 0,551416\ CF_{t-4} + 0,413562\ CF_{t-5}$$
$$+ 0,27508\ CF_{t-6} + 0,137854\ CF_{t-7} + e_t$$

- Se observa que la estimación del parámetro S es -0.001831 para lo que es ZIED. Entonces tenemos:

$$\delta_i = \begin{cases} (7+1-i)\delta & 0 \le i \le 7 \\ 0 & i > 7 \end{cases} = \{8\delta, 7\delta, 6\delta, 5\delta, 4\delta, 3\delta, 2\delta, 1\delta\} =$$

$$\{-0,014648;\ -0,012817;\ -0,010986;\ -0,009155;\ -0,007324;\ -0,005493;\ -0,003662;\ -0,001831\}$$

Por lo tanto, la estimación del modelo inicial será la siguiente:

$$Y_t = -38253,68 - 0,014648\ IED_t - 0,012817\ IED_{t-1} - 0,010986\ IED_{t-2}$$
$$- 0,009155\ IED_{t-3} - 0,007324\ IED_{t-4} - 0,005493\ IED_{t-5}$$
$$- 0,003662\ IED_{t-6} - 0,001831\ IED_{t-7} + e_t$$

## 1.- Heterocedasticidad

**Prueba de Glejser.**

Se escoge el supuesto para verificar la ausencia o presencia de heterocedasticidad en nuestro modelo mediante el test de Glejser.

**Gráfico 17. Test de Heterocedasticidad**

Heteroskedasticity Test: Glejser

| | | | |
|---|---|---|---|
| F-statistic | 0.681293 | Prob. F(2,49) | 0.5107 |
| Obs*R-squared | 1.406886 | Prob. Chi-Square(2) | 0.4949 |
| Scaled explained SS | 1.606247 | Prob. Chi-Square(2) | 0.4479 |

Test Equation:
Dependent Variable: ARESID
Method: Least Squares
Date: 08/01/23   Time: 15:19
Sample: 2010Q2 2023Q1
Included observations: 52

| Variable | Coefficient | Std. Error | t-Statistic | Prob. |
|---|---|---|---|---|
| C | -259.9107 | 3101.230 | -0.083809 | 0.9335 |
| ZCF | 0.004138 | 0.009175 | 0.451018 | 0.6540 |
| ZIED | 0.000271 | 0.000490 | 0.553458 | 0.5825 |

| | | | |
|---|---|---|---|
| R-squared | 0.027056 | Mean dependent var | 1754.287 |
| Adjusted R-squared | -0.012657 | S.D. dependent var | 1631.668 |
| S.E. of regression | 1641.961 | Akaike info criterion | 17.70113 |
| Sum squared resid | 1.32E+08 | Schwarz criterion | 17.81370 |
| Log likelihood | -457.2294 | Hannan-Quinn criter. | 17.74429 |
| F-statistic | 0.681293 | Durbin-Watson stat | 2.205232 |
| Prob(F-statistic) | 0.510690 | | |

*Fuente: Los Autores, 2023.*

La probabilidad de F es mayor al 5%, por ende, se confirma que el modelo no tiene problemas de heterocedasticidad, por ende, no es necesaria su corrección

## 2.- Normalidad

Se verifica la ausencia o presencia de normalidad en nuestro modelo mediante el análisis de Jarque-Bera.

**Gráfico 18. Prueba de Normalidad Jarque-Bera**

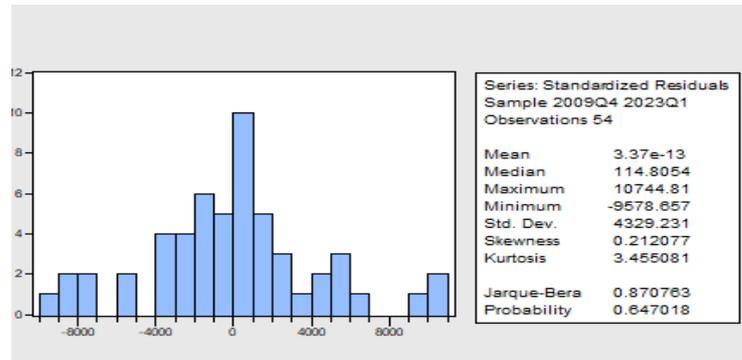

*Fuente: Los Autores, 2023.*

Si es mayor al 5% hay presencia de normalidad.

El resultado de Jarque-Bera en la tabla es 0.6470

En el caso de nuestro modelo, la probabilidad de Jarque-Bera es mayor al 5%, por lo tanto, nuestro modelo tiene normalidad en los residuos.

## 4.- Linealidad

Se verifica la ausencia o presencia de linealidad en nuestro modelo mediante el análisis de Ramsey RESET.

**Gráfico 19. Test de Linealidad Ramsey Reset**

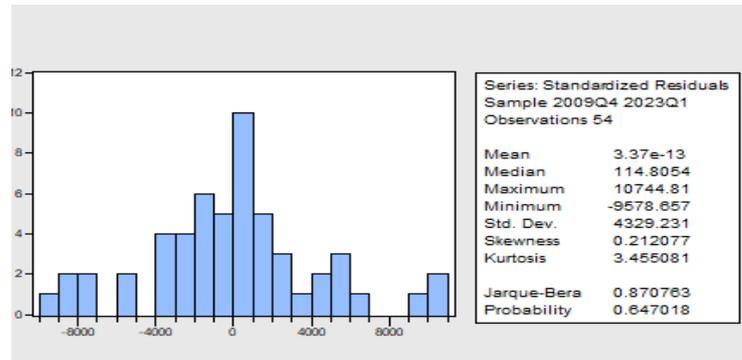

*Fuente: Los Autores, 2023.*

Si la probabilidad es menor al 5% existe problema de linealidad y al ser mayor al 5%, no existe.

El resultado de F en la tabla es 0.0059

La probabilidad del F estadístico es menor al 5%, podemos afirmar que no existe problema de linealidad

**5.- Autocorrelación**

Se verifica la ausencia o presencia de autocorrelación en nuestro modelo mediante el análisis de estadístico Durbin-Watson.

**Gráfico 20. Test de autocorrelación corregida.**

Equation: EQ01   Workfile: PROYECTO2::Untitled\

View | Proc | Object | Print | Name | Freeze | Estimate | Forecast | Stats | Resids

Dependent Variable: PIB
Method: Least Squares
Date: 08/01/23  Time: 15:04
Sample (adjusted): 2010Q2 2023Q1
Included observations: 52 after adjustments
Convergence achieved after 85 iterations

| Variable | Coefficient | Std. Error | t-Statistic | Prob. |
|---|---|---|---|---|
| C | -38253.68 | 119957.6 | -0.318893 | 0.7512 |
| ZCF | 0.137854 | 0.039814 | 3.462434 | 0.0012 |
| ZIED | -0.001831 | 0.013322 | -0.137439 | 0.8913 |
| AR(1) | 0.793639 | 0.145889 | 5.440025 | 0.0000 |
| AR(2) | 0.226586 | 0.155021 | 1.461653 | 0.1505 |
| R-squared | 0.956823 | Mean dependent var | | 44004.71 |
| Adjusted R-squared | 0.953148 | S.D. dependent var | | 11590.27 |
| S.E. of regression | 2508.753 | Akaike info criterion | | 18.58417 |
| Sum squared resid | 2.96E+08 | Schwarz criterion | | 18.77179 |
| Log likelihood | -478.1884 | Hannan-Quinn criter. | | 18.65610 |
| F-statistic | 260.3833 | Durbin-Watson stat | | 1.997213 |
| Prob(F-statistic) | 0.000000 | | | |

| Inverted AR Roots | 1.02 | -.22 | | |
| | Estimated AR process is nonstationary | | | |

*Fuente: Los Autores, 2023.*

Podemos visualizar que DW se encuentra en 1,99 %. Por ende, podemos afirmar no hay problema de autocorrelación en nuestro modelo.

**Anexo 5. Modelo de Vectores Autorregresivos**

Se usan 7 rezagos dentro de la opción as VAR.

**Gráfico 45. Pasos para aplicación del VAR**

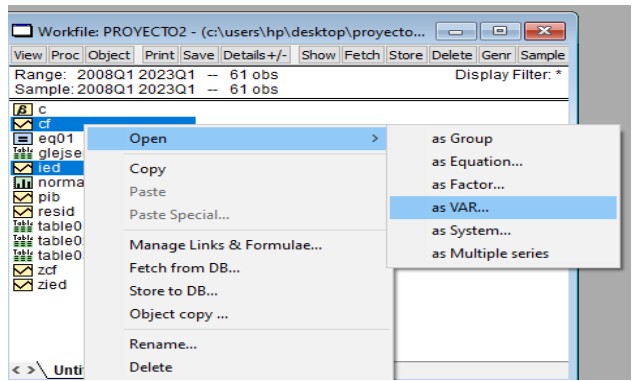

*Fuente: Los Autores, 2023.*

**Gráfico 46. aplicación del VAR con 7 rezagos**

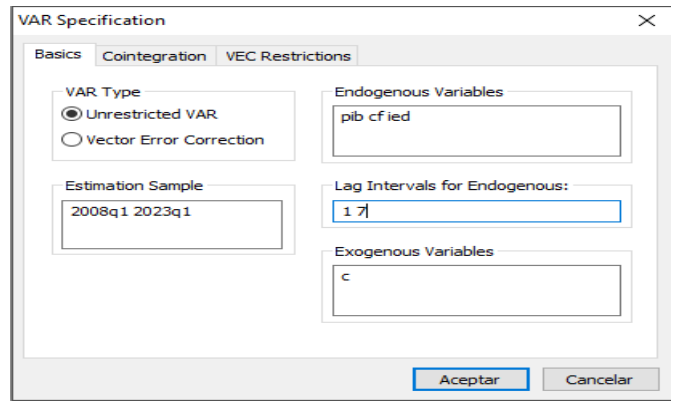

*Fuente: Los Autores, 2023.*

**Tabla 1. Modelo VAR**

Vector Autoregression Estimates
Date: 08/01/23   Time: 19:47
Sample (adjusted): 2009Q2 2023Q1
Included observations: 56 after adjustments
Standard errors in ( ) & t-statistics in [ ]

|  | PIB | CF | IED |
|---|---|---|---|
| PIB(-1) | 0.469584 | -0.106175 | -0.076451 |
|  | (0.19452) | (0.07307) | (0.20156) |
|  | [ 2.41411] | [-1.45298] | [-0.37930] |
| PIB(-2) | 0.577223 | 0.160100 | 0.367789 |
|  | (0.21624) | (0.08124) | (0.22407) |
|  | [ 2.66930] | [ 1.97079] | [ 1.64138] |
| PIB(-3) | 0.070912 | 0.085277 | 0.197547 |
|  | (0.23923) | (0.08987) | (0.24789) |
|  | [ 0.29643] | [ 0.94890] | [ 0.79693] |
| PIB(-4) | 0.533092 | 0.006404 | -0.541997 |
|  | (0.23621) | (0.08874) | (0.24476) |
|  | [ 2.25686] | [ 0.07217] | [-2.21439] |
| PIB(-5) | -0.524897 | -0.139985 | -0.180093 |
| | (0.18859) | (0.07085) | (0.19541) |
| | [-2.78335] | [-1.97592] | [-0.92160] |
| | | | |
| CF(-1) | 0.409645 | 1.005883 | 0.805520 |
| | (0.48497) | (0.18219) | (0.50252) |
| | [ 0.84469] | [ 5.52117] | [ 1.60295] |
| | | | |
| CF(-2) | -1.592630 | -0.399548 | -0.527640 |
| | (0.63898) | (0.24004) | (0.66211) |
| | [-2.49247] | [-1.66448] | [-0.79691] |
| | | | |
| CF(-3) | -0.619526 | -0.190275 | -0.273190 |
| | (0.69815) | (0.26227) | (0.72342) |
| | [-0.88739] | [-0.72548] | [-0.37764] |
| | | | |
| CF(-4) | 0.685952 | 0.592633 | 1.667247 |
| | (0.70721) | (0.26568) | (0.73281) |
| | [ 0.96995] | [ 2.23067] | [ 2.27515] |
| | | | |
| CF(-5) | 0.308967 | -0.248007 | 0.102626 |
| | (0.54832) | (0.20599) | (0.56817) |
| | [ 0.56348] | [-1.20400] | [ 0.18063] |
| | | | |
| IED(-1) | 0.056520 | -0.031755 | 0.802928 |
| | (0.15485) | (0.05817) | (0.16045) |
| | [ 0.36501] | [-0.54589] | [ 5.00418] |
| | | | |
| IED(-2) | 0.127005 | 0.083757 | 0.118787 |
| | (0.20024) | (0.07523) | (0.20749) |
| | [ 0.63425] | [ 1.11341] | [ 0.57248] |
| | | | |
| IED(-3) | -0.067690 | 0.004149 | 0.146091 |
| | (0.20306) | (0.07629) | (0.21042) |
| | [-0.33334] | [ 0.05439] | [ 0.69430] |
| | | | |
| IED(-4) | -0.118142 | -0.125452 | -0.487628 |
| | (0.20580) | (0.07731) | (0.21325) |
| | [-0.57407] | [-1.62267] | [-2.28666] |
| | | | |
| IED(-5) | -0.018828 | 0.070208 | 0.413132 |
| | (0.15795) | (0.05934) | (0.16367) |
| | [-0.11920] | [ 1.18318] | [ 2.52414] |
| | | | |
| C | 4878.744 | 2173.394 | -8370.813 |
| | (3009.67) | (1130.64) | (3118.64) |
| | [ 1.62102] | [ 1.92227] | [-2.68413] |
| R-squared | 0.989342 | 0.864282 | 0.994400 |
| Adj. R-squared | 0.985345 | 0.813388 | 0.992300 |
| Sum sq. resids | 87950011 | 12412151 | 94433491 |
| S.E. equation | 1482.818 | 557.0492 | 1536.502 |
| F-statistic | 247.5269 | 16.98195 | 473.5050 |
| Log likelihood | -478.9345 | -424.1079 | -480.9261 |
| Akaike AIC | 17.67623 | 15.71814 | 17.74736 |
| Schwarz SC | 18.25491 | 16.29681 | 18.32603 |
| Mean dependent | 42621.81 | 10626.30 | 43318.22 |
| S.D. dependent | 12248.71 | 1289.507 | 17509.71 |

| | |
|---|---|
| Determinant resid covariance (dof adj.) | 8.64E+17 |
| Determinant resid covariance | 3.15E+17 |
| Log likelihood | -1366.544 |
| Akaike information criterion | 50.51943 |
| Schwarz criterion | 52.25544 |
| Number of coefficients | 48 |

*Fuente: Los Autores, 2023.*

**Gráfico 47. Pasos para Lag Structure**

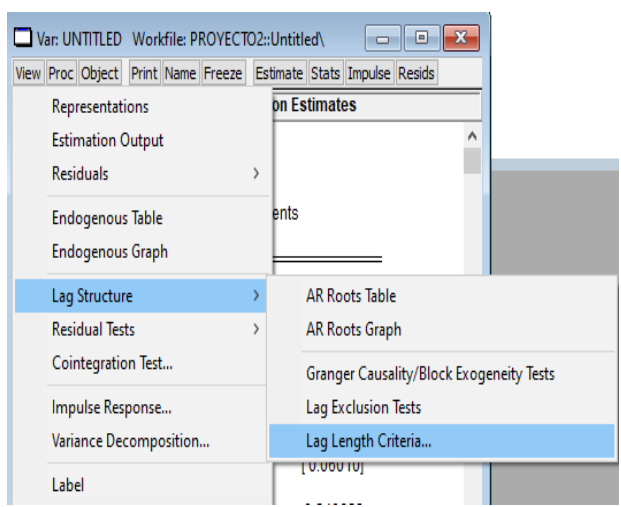

*Fuente: Los Autores, 2023.*

**Gráfico 48. VAR Lar Order Selection Criteria**

VAR Lag Order Selection Criteria
Endogenous variables: PIB CF IED
Exogenous variables: C
Date: 08/01/23   Time: 19:45
Sample: 2008Q1 2023Q1
Included observations: 54

| Lag | LogL | LR | FPE | AIC | SC | HQ |
|---|---|---|---|---|---|---|
| 0 | -1601.888 | NA | 1.31e+22 | 59.44031 | 59.55081 | 59.48292 |
| 1 | -1381.165 | 408.7473 | 5.15e+18 | 51.59870 | 52.04069* | 51.76916 |
| 2 | -1365.724 | 26.87816 | 4.07e+18 | 51.36015 | 52.13365 | 51.65846 |
| 3 | -1357.036 | 14.15771 | 4.16e+18 | 51.37172 | 52.47671 | 51.79787 |
| 4 | -1336.308 | 31.47597 | 2.73e+18 | 50.93735 | 52.37383 | 51.49134 |
| 5 | -1316.976 | 27.20857* | 1.91e+18* | 50.55466* | 52.32265 | 51.23651* |
| 6 | -1312.868 | 5.325495 | 2.39e+18 | 50.73584 | 52.83532 | 51.54553 |
| 7 | -1304.095 | 10.39679 | 2.55e+18 | 50.74427 | 53.17526 | 51.68181 |

*Fuente: Los Autores, 2023.*

Nos fijamos en los * que son el criterio de selección para la orden de Var que en este caso es de orden 5

**Gráfico 49. aplicación del VAR con 7 rezagos**

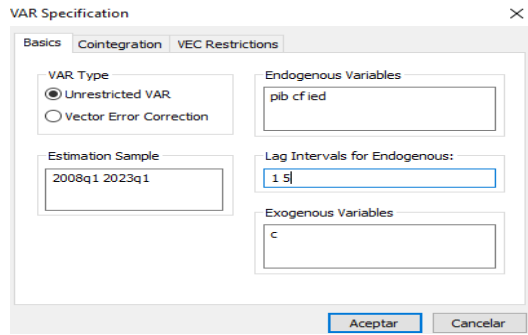

*Fuente: Los Autores, 2023.*

**Gráfico 49. Tabla de VAR – parte 1.**

| Vector Autoregression Estimates | | | |
|---|---|---|---|
| Date: 08/01/23 Time: 19:47 | | | |
| Sample (adjusted): 2009Q2 2023Q1 | | | |
| Included observations: 56 after adjustments | | | |
| Standard errors in ( ) & t-statistics in [ ] | | | |
| | PIB | CF | IED |
| PIB(-1) | 0.469584 | -0.106175 | -0.076451 |
| | (0.19452) | (0.07307) | (0.20156) |
| | [ 2.41411] | [-1.45298] | [-0.37930] |
| PIB(-2) | 0.577223 | 0.160100 | 0.367789 |
| | (0.21624) | (0.08124) | (0.22407) |
| | [ 2.66930] | [ 1.97079] | [ 1.64138] |
| PIB(-3) | 0.070912 | 0.085277 | 0.197547 |
| | (0.23923) | (0.08987) | (0.24789) |
| | [ 0.29643] | [ 0.94890] | [ 0.79693] |
| PIB(-4) | 0.533092 | 0.006404 | -0.541997 |
| | (0.23621) | (0.08874) | (0.24476) |
| | [ 2.25686] | [ 0.07217] | [-2.21439] |
| PIB(-5) | -0.524897 | -0.139985 | -0.180093 |
| | (0.18859) | (0.07085) | (0.19541) |
| | [-2.78335] | [-1.97592] | [-0.92160] |
| CF(-1) | 0.409645 | 1.005883 | 0.805520 |
| | (0.48497) | (0.18219) | (0.50252) |
| | [ 0.84469] | [ 5.52117] | [ 1.60295] |
| CF(-2) | -1.592630 | -0.399548 | -0.527640 |
| | (0.63898) | (0.24004) | (0.66211) |
| | [-2.49247] | [-1.66448] | [-0.79691] |
| CF(-3) | -0.619526 | -0.190275 | -0.273190 |
| | (0.69815) | (0.26227) | (0.72342) |
| | [-0.88739] | [-0.72548] | [-0.37764] |

*Fuente: Los Autores, 2023.*

**Gráfico 50. Tabla de VAR – parte 2.**

| | Vector Autoregression Estimates | | |
|---|---|---|---|
| CF(-4) | 0.685952 | 0.592633 | 1.667247 |
| | (0.70721) | (0.26568) | (0.73281) |
| | [ 0.96995] | [ 2.23067] | [ 2.27515] |
| CF(-5) | 0.308967 | -0.248007 | 0.102626 |
| | (0.54832) | (0.20599) | (0.56817) |
| | [ 0.56348] | [-1.20400] | [ 0.18063] |
| IED(-1) | 0.056520 | -0.031755 | 0.802928 |
| | (0.15485) | (0.05817) | (0.16045) |
| | [ 0.36501] | [-0.54589] | [ 5.00418] |
| IED(-2) | 0.127005 | 0.083757 | 0.118787 |
| | (0.20024) | (0.07523) | (0.20749) |
| | [ 0.63425] | [ 1.11341] | [ 0.57248] |
| IED(-3) | -0.067690 | 0.004149 | 0.146091 |
| | (0.20306) | (0.07629) | (0.21042) |
| | [-0.33334] | [ 0.05439] | [ 0.69430] |
| IED(-4) | -0.118142 | -0.125452 | -0.487628 |
| | (0.20580) | (0.07731) | (0.21325) |
| | [-0.57407] | [-1.62267] | [-2.28666] |
| IED(-5) | -0.018828 | 0.070208 | 0.413132 |
| | (0.15795) | (0.05934) | (0.16367) |
| | [-0.11920] | [ 1.18318] | [ 2.52414] |
| C | 4878.744 | 2173.394 | -8370.813 |
| | (3009.67) | (1130.64) | (3118.64) |
| | [ 1.62102] | [ 1.92227] | [-2.68413] |
| R-squared | 0.989342 | 0.864282 | 0.994400 |
| Adj. R-squared | 0.985345 | 0.813388 | 0.992300 |
| Sum sq. resids | 87950011 | 12412151 | 94433491 |
| S.E. equation | 1482.818 | 557.0492 | 1536.502 |
| F-statistic | 247.5269 | 16.98195 | 473.5050 |
| Log likelihood | -478.9345 | -424.1079 | -480.9261 |
| Akaike AIC | 17.67623 | 15.71814 | 17.74736 |

*Fuente: Los Autores, 2023.*

**Gráfico 50. Representación del modelo VAR.**

*Fuente: Los Autores, 2023.*

## Las ecuaciones nos quedan de la siguiente manera:

VAR Model - Substituted Coefficients:

================================

$$PIB = 0.469583869954 * PIB(-1) + 0.577222710481 * PIB(-2) + 0.0709124520854$$
$$* PIB(-3) + 0.533091813254 * PIB(-4) - 0.524897417444 * PIB(-5)$$
$$+ 0.409644724565 * CF(-1) - 1.59262975087 * CF(-2) - 0.619526199112$$
$$* CF(-3) + 0.68595183443 * CF(-4) + 0.308966878161 * CF(-5)$$
$$+ 0.0565198428591 * IED(-1) + 0.127005337163 * IED(-2)$$
$$- 0.0676897401283 * IED(-3) - 0.118142178173 * IED(-4)$$
$$- 0.0188281161283 * IED(-5) + 4878.74389983$$

$$CF = -0.106174661476 * PIB(-1) + 0.160099918819 * PIB(-2) + 0.085277223556$$
$$* PIB(-3) + 0.00640420640503 * PIB(-4) - 0.139985447631 * PIB(-5)$$
$$+ 1.00588347822 * CF(-1) - 0.39954754107 * CF(-2) - 0.190274579136$$
$$* CF(-3) + 0.592633017683 * CF(-4) - 0.248006550055 * CF(-5)$$
$$- 0.0317549750057 * IED(-1) + 0.0837566638227 * IED(-2)$$
$$+ 0.00414880340431 * IED(-3) - 0.125452088498 * IED(-4)$$
$$+ 0.0702082434019 * IED(-5) + 2173.39432447$$

$$IED = -0.0764505545951 * PIB(-1) + 0.367789024292 * PIB(-2) + 0.197547277242$$
$$* PIB(-3) - 0.541996650472 * PIB(-4) - 0.180092869797 * PIB(-5)$$
$$+ 0.805520458105 * CF(-1) - 0.52764041198 * CF(-2) - 0.273190061991$$
$$* CF(-3) + 1.66724715955 * CF(-4) + 0.102625852822 * CF(-5)$$
$$+ 0.802928438988 * IED(-1) + 0.118786579415 * IED(-2)$$
$$+ 0.146091053293 * IED(-3) - 0.487627767138 * IED(-4)$$
$$+ 0.413132293929 * IED(-5) - 8370.81309579$$

Procedemos a verificar a traves del grafico en Lag Structure

**Gráfico 50. Pasos para la prueba de estabilidad.**

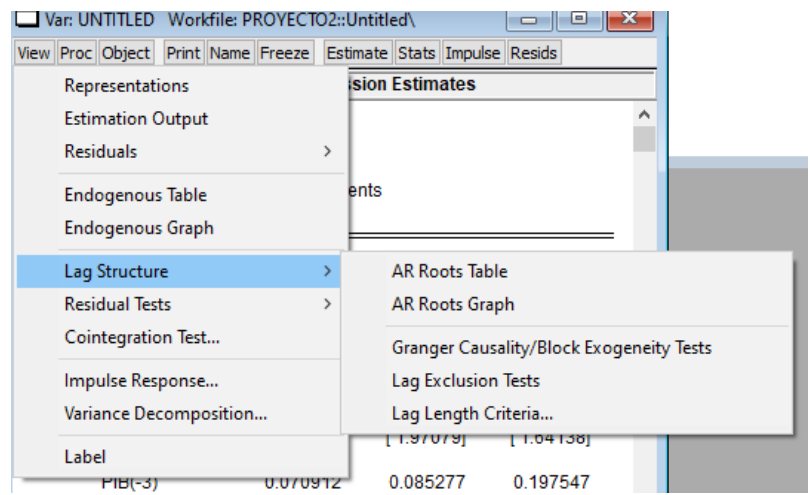

*Fuente: Los Autores, 2023.*

**Gráfico 21. Prueba de estabilidad**

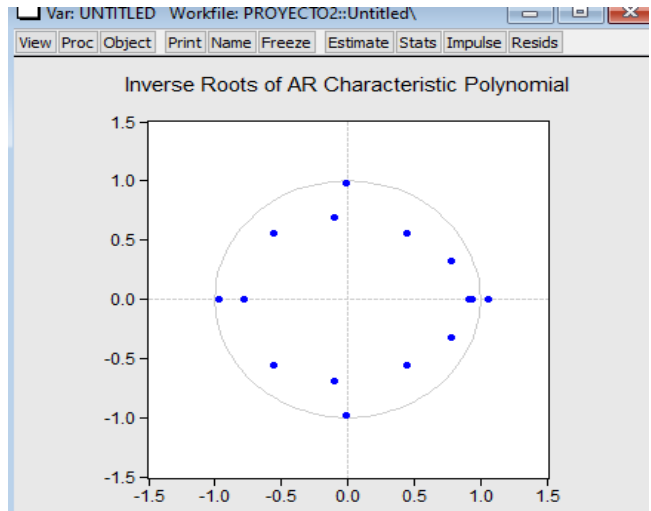

*Fuente: Los Autores, 2023.*

La regresión es estable.